%% file: main.tex
\documentclass[fleqn,usenatbib,useAMS]{mnras}

\usepackage{graphicx}	
\usepackage{amsmath}	
\usepackage{amssymb}	
\usepackage{multicol}        
\usepackage{bm}		
\usepackage{pdflscape}	
\usepackage{float} 
\usepackage{multirow} 

\newcommand{\Tab}[1]{Table~\ref{#1}}
\newcommand{\Sec}[1]{\S\ref{#1}}
\newcommand{\Sub}[1]{\S\S\ref{#1}}

\newcommand{\Eq}[1]{equation~(\ref{#1})}
\newcommand{\Fig}[1]{Fig.~\ref{#1}}

\newcommand{\rhocoll}{$\rho_{c}(z_\mathrm{-2})$}
\newcommand{\Menc}{$ M_{-2}\,$}

\usepackage[T1]{fontenc}
\usepackage{ae,aecompl}

\usepackage{newtxtext,newtxmath}

\title[The cosmology dependence of the concentration-mass-redshift relation]{The cosmology dependence of the concentration-mass-redshift relation}

\author[L\'{o}pez-Cano et al.]
{Daniel L\'{o}pez-Cano,$^{1,2}$\thanks{Contact e-mail: \href{mailto:daniellopezcano13@gmail.com}{daniellopezcano13@gmail.com}}
Ra\'{u}l E. Angulo,$^{1,3}$
Aaron D. Ludlow,$^{4}$
M. Zennaro,$^{1}$
S. Contreras,$^{1}$
\newauthor
Jonás Chaves-Montero,$^{1}$
G. Aricò,$^{1,5}$
\\
$^{1}$Donostia International Physics Center (DIPC), Paseo Manuel de Lardizabal, 4, 20018 Donostia-San Sebastián, Spain\\
$^{2}$Departamento de F\'{\i}sica Te\'{o}rica, M\'{o}dulo 15, Facultad de Ciencias, Universidad Aut\'{o}noma de Madrid, 28049 Madrid, Spain\\
$^{3}$IKERBASQUE, Basque Foundation for Science, 48013, Bilbao, Spain.\\
$^{4}$International Centre for Radio Astronomy Research, University of Western Australia, 35 Stirling Highway, Crawley, Western Australia, 6009, Australia\\
$^{5}$Institute for Computational Science, University of Zurich, Winterthurerstrasse 190, 8057 Zurich, Switzerland.
}

\date{Last updated 2021 March 16; in original form 2021 March 16}

\pubyear{2022}

\begin{document}
\label{firstpage}
\pagerange{\pageref{firstpage}--\pageref{lastpage}}
\maketitle

\begin{abstract}
The concentrations of dark matter haloes provide crucial information about their internal structure and how it depends on mass and redshift -- the so-called concentration-mass-redshift relation, denoted $c(M,z)$. We present here an extensive study of the cosmology-dependence of $c(M,z)$ that is based on a suite of 72 gravity-only, full N-body simulations in which the following cosmological parameters were varied: $\sigma_{8}$, $\Omega_{\mathrm{M}}$, $\Omega_{\mathrm{b}}$, $n_{\mathrm{s}}$, $h$, $M_{\nu}$, $w_{0}$ and $w_{\mathrm{a}}$. We characterize the impact of these parameters on concentrations for different halo masses and redshifts. In agreement with previous works, and for all cosmologies studied, we find that there exists a tight correlation between the characteristic densities of dark matter haloes within  their scale radii, $r_{-2}$, and the critical density of the Universe at a suitably defined formation time. This finding, when combined with excursion set modelling of halo formation histories, allows us to accurately predict the concentrations of dark matter haloes as a function of mass, redshift, and cosmology. We use our simulations to test the reliability of a number of published models for predicting halo concentration and highlight when they succeed or fail to reproduce the cosmological $c(M,z)$ relation.
\end{abstract}

\begin{keywords}
  methods: numerical, statistical -- cosmology: theory, dark matter
\end{keywords}


\input{sec1}

\input{sec2}

\input{sec3}

\input{sec4}

\input{sec5}


\section*{Acknowledgements}

\addcontentsline{toc}{section}{Acknowledgements}
DLC, REA, MZ, and SC acknowledge the support of the ERC-StG number 716151 (BACCO). ADL acknowledges financial support from the Australian Research Council through their Future Fellowship scheme (project number FT160100250). SC acknowledges the support of the “Juan de la Cierva Incorporaci\'on” fellowship (IJC2020-045705-I). JCM acknowledges partial support from the Spanish Ministry of Science, Innovation and Universities (MCIU/AEI/FEDER, UE) through the grant PGC2018-097585-B-C22. The authors acknowledge the computer resources at MareNostrum and the technical support provided by Barcelona Supercomputing Center (RES-AECT-2019-2-0012, RES-AECT-2020-3-0014). All the authors acknowledge the assistance of Matteo Esposito and Marcos Pellejero-Ibañez for their help running the cosmological simulations. DLC further thanks Benedict Diemer, Jens Stücker, Marcos Pellejero-Ibañez and Daniel Muñoz-Segovia for their insightful discussions. This work made use of the following software: \textsc{Python}, \textsc{Matplotlib} \citep{Matplotlib-Hunter07} and \textsc{Numpy} \citep{Numpy-vanDerWalt11}.\\

DLC generated the software necessary for the analysis of the cosmological simulations (which he helped to run), obtained, validated and interpreted the results, and wrote the original and final versions of the article. REA and ADL designed the project and contributed to the interpretation of results and writing of the paper. MZ provided technical tools and participated in discussions for the interpretation of the results. SC run most of the simulations used in this paper, help to run the merger trees and contribute to the codes used to analyse simulations. JCM contributed to the interpretation of the results and helped to develop the software for the merger tree analysis. GA provided useful resources for developing the software for the density profile analysis and helped editing the final version of the article.


\bibliographystyle{mnras}
\bibliography{archive}

\appendix

\input{A1}

\input{A2}

\bsp	
\label{lastpage}
\end{document}

%% file: sec1.tex
\section{Introduction}\label{sec1}
Cosmological simulations have revealed that the spherically-averaged density profiles of dark matter (DM) haloes exhibit a high degree of self-similarity across a wide range of masses, redshifts, and cosmologies (\citealt{NFW, 1999ApJ...517...64H, 2001ApJ...556...93B, 2001MNRAS.321..559B, 2007MNRAS.381.1450N,2008MNRAS.391.1940M,  2008MNRAS.385..545K, 2009MNRAS.396..709W, 2013JCAP...10..012H, 2017MNRAS.465L..84L, 2020MNRAS.495.4994B} and \citealt{AnguloHahn2022} for a review). The most popular analytic expression used to describe these profiles is the NFW profile \citep{NFW, 1997ApJ...490..493N}, written as
\begin{equation}\label{eq_NFW}
    \rho_{\textrm{NFW}} = \frac{4\,\rho_{-2}}{r/r_{-2}\left(r/r_{-2}+1\right)^2},
\end{equation}
where $r_{-2}$ is the characteristic radius at which the profile's logarithmic slope is equal to $-2$, and $\rho_{-2}=\rho_{\rm NFW}(r_{-2})$ is the corresponding density. These two parameters fully specify the NFW profile and therefore completely describe the structure of dark matter haloes. It is however common practice to recast these parameters in terms of the halo's virial mass\footnote{We define the virial mass $M_{200,\mathrm{m}}$ of a DM halo as the total mass enclosed by a sphere of radius $r_{200,\mathrm{m}}$, centered on the halo particle with the minimum potential energy, that encloses a mean density of $200\times\rho_{\rm m}$, where $\rho_\textrm{m} = \Omega_{\mathrm{m}}\rho_{\mathrm{c}}$ is the mean matter density and $\rho_{\rm c} = 3\,H^2/8\pi G$ is the critical density of the universe.}, $M_{200,m}$, and concentration, $c=r_{200,m}/r_{-2}$, an approach we follow in this paper; in what follows we refer to these quantities simply as $M_{200}$ and $r_{200}$.

As simulations grew in volume and simultaneously achieved higher mass and spatial resolution, it became clear that simulated halo profiles exhibit slight but systematic departures from the NFW shape. As discussed in \citet[][see also \citealt{2008MNRAS.387..536G,2011MNRAS.415.3895L,2014MNRAS.441.3359D,  2018ApJ...859...55C}]{2004MNRAS.349.1039N}, simulated halo profiles are better described by the \citet{1965TrAlm...5...87E} profile, which can be written
\begin{equation}
    \label{eq_Einasto}
    \rho_{\textrm{E}}= \rho_{-2}\exp\left(-\frac{2}{\alpha}\left[\left(\frac{r}{r_{-2}}\right)^\alpha-1\right]\right),
\end{equation}
where $r_{-2}$ and $\rho_{-2}$ have the same meaning as in \Eq{eq_NFW}, and $\alpha$ is a shape parameter that can be tailored to better-fit individual haloes. For $\alpha \approx 0.18$, \Eq{eq_Einasto} resembles the NFW profile over a wide range of scales.

Neglecting the slight deviations between simulated halo density profiles and the NFW profile, the values of $c$ and $M_{\rm 200}$ are sufficient to determine their structure. This led to numerous studies of the relationship between halo mass and concentration, and how it changes as a function of redshift and cosmology (the so-called concentration-mass-redshift relation, often denoted $c(M,z)$). These studies paint a clear picture of the structure of CDM haloes: at fixed redshift, their concentrations, on average, decrease with increasing mass, and at fixed mass, on average, decrease with increasing redshift \citep[e.g.][]{2001MNRAS.321..559B,2012MNRAS.423.3018P,2012MNRAS.427.1322L,2013MNRAS.432.1103L,2014MNRAS.441..378L, 2015MNRAS.452.1217C, 2016MNRAS.460.1214L, 2019ApJ...871..168D, 2020MNRAS.495.4994B}. Although the exact physical mechanism that sets the concentration of a halo is not known, numerous studies have convincingly demonstrated that it is closely connected to its assembly history \citep{NFW, 1997ApJ...490..493N, 2001MNRAS.321..559B, 2002ApJ...568...52W, 2003ApJ...597L...9Z, Ludlow_2014, 2016MNRAS.460.1214L, 2019ApJ...871..168D}.

A number of studies have also addressed the mass and redshift dependence of $\alpha$, the Einasto shape parameter in \Eq{eq_Einasto}. For example, \citet[][see also \citealt{2014MNRAS.441.3359D,2018ApJ...859...55C}]{2008MNRAS.387..536G} demonstrated that the average value of $\alpha$ increases with both halo mass and redshift in a manner that can be neatly described by a single relation between $\alpha$ and peak height\footnote{The peak height, a dimensionless mass parameter, is defined as $\nu(M,z) = \delta_{\mathrm{c}}/\sigma(M,z)$, where $\sigma(M,z)$ is the variance of the matter density perturbations linearly extrapolated to redshift $z$, and $\delta_\mathrm{c}$ is the critcal density for gravitational collapse, usually estimated from the spherical collapse model for which $\delta_{\rm c}\approx 1.686$ \citep[e.g.][]{1980lssu.book.....P}.}, $\nu(M,z)$. \citet[][see also \citealt{2017MNRAS.465L..84L}]{2013MNRAS.432.1103L} showed that, like the halo concentration, $\alpha$ is intimately linked to the assembly histories of dark matter haloes. 

Given the approximate self-similarity of halo structure, the ability to accurately predict halo concentrations has numerous applications, including estimating merger rates of primordial black holes \citep[e.g.][]{2016PhRvL.117t1102M}, predicting the lensing signal associated with haloes \citep[e.g.][]{2002A&A...396...21B, 2007A&A...473..715F,  2008JCAP...08..006M, 2021arXiv210900018A} and their substructure \citep[e.g.][]{10.1093/mnras/sty159}, and to estimate the gamma ray signal potentially produced by dark matter annihilation \citep[e.g.][]{2014MNRAS.442.2271S, 2018JCAP...08..019O}. Another potential application -- indeed, the one that motivated this work -- is to improve the performance of cosmological rescaling algorithms \citep{10.1111/j.1365-2966.2010.16459.x, 2020MNRAS.499.4905C} that can be used to transmute a template N-body simulation carried out with a set of cosmological parameters to a synthetic simulation consistent with another cosmology. Whether existing models can appropriately account for the cosmology dependence of halo concentrations has not been rigorously tested. 

The aim of this work is therefore to study the dependence of the $c(M,z)$ relation on cosmology, and to test the extent to which it can be reproduced by published models for predicting halo concentrations. To do so, we ran a large suite of gravity-only simulations in which the cosmological parameters were systematically varied with respect to the best-fit \cite{2020A&A...641A...6P} results. In \Sec{sec2} we present our suite of cosmological simulations along with their associated halo and merger tree catalogs  (\Sub{subsec_simulations}), explain our approach to discarding unrelaxed haloes (\Sub{subsec_relaxedness}), and outline how we measure halo concentrations (\Sub{subsec_obtain_rs}). In \Sec{sec3} we present the $c(M,z)$ relations obtained for different cosmologies (\Sub{subsec_c_M_relation}) and study the relation between the internal structure of haloes and their formation histories. In \Sub{subsec_c_vs_theta} we compare the performance of different published models for predicting the mass- and redshift-dependence of halo concentration, focusing on their ability to reproduce the cosmology-dependence of the $c(M,z)$ relation. In \Sec{sec4} we discuss how accurate predictions for halo concentration can lead to improved accuracy when applied to a cosmological scaling algorithm. In \Sec{sec5} we provide a few concluding remarks.

%% file: sec2.tex
\section{Numerical simulations and analysis}\label{sec2}

Below we describe the pertinent 
details of the numerical simulation used in this work, and discuss our analysis algorithms and techniques.

\subsection{Numerical simulations}\label{subsec_simulations}

Our results are inferred from a suite of DM-only simulations in which we modify the values of different cosmological parameters. We define these parameters below.

\begin{enumerate}
    \item $\sigma_{8}$: The root mean square of matter density perturbations averaged in spheres of radius $R=8\textit{h}^{-1}\textrm{Mpc}$ and linearly extrapolated to $z=0$.\\
    
    \item $\Omega_{\mathrm{m}}$: The dimensionless matter density parameter, $\Omega_{\mathrm{m}}\equiv\rho_{\mathrm{m}}/\rho_\mathrm{c}=8\pi G \rho_{\mathrm{m}} / 3H^2$, which is the ratio of the total matter density, $\rho_\textrm{m}$, and the critical density, $\rho_{\rm c}$. Note that $\Omega_{\rm m}$ includes contributions from both DM and baryons, i.e. $\Omega_{\rm m}=\Omega_{\rm cdm}+\Omega_{\rm b}$. For runs in which $\Omega_{\mathrm{m}}$ is varied, we only modify the value of $\Omega_{\mathrm{cdm}}$ (keeping $\Omega_{\mathrm{b}}$ fixed) and adjust the value of $\Omega_{\textrm{DE}}$ (the cosmic dark energy density) to maintain a  flat cosmology. Note that neutrinos do not contribute to this definition of $\Omega_{\mathrm{m}}$.\\
    
    \item $n_{\mathrm{s}}$: The scalar spectral index of the primordial density fluctuation power spectrum, $P(k) \propto k^{n_{\mathrm{s}}-1}$.\\

    \item $w_{0}$ and $w_\textrm{a}$: The dynamical dark energy parameters used in the Chevallier-Polarski-Linder (CPL) parameterization~\citep{2001IJMPD..10..213C, 2003PhRvL..90i1301L}. When $w_{0}=-1$ and $w_\textrm{a}=0$ the dark energy contribution to the background expansion is consistent with a cosmological constant, see \Eq{eq_friedmann}.\\
    
\item $M_{\nu}$: The sum of the individual masses for the three neutrino species, which is related to the neutrino density parameter by $\Omega_\nu = M_\nu / [(93.14 \textrm{eV}) h^2]$ (with $M_{\nu}$ expressed in $\textrm{eV}$)\footnote{The first Friedmann equation can be written in terms of the neutrino density parameter, $\Omega_\nu$, as~\citep{2017MNRAS.466.3244Z}:
    \begin{align}\label{eq_friedmann}
H^{2}(a)=H_{0}^{2} &\left[\left(\Omega_{\mathrm{cdm}, 0}+\Omega_{\mathrm{b}, 0}\right) a^{-3}+\Omega_{v}(a) E^{2}(a)+\right.\nonumber\\
&\quad\quad\quad\quad\quad\quad\quad\quad\quad\left.+\Omega_{\mathrm{DE}, 0} a^{-3\left(1+w_{0}+w_{\mathrm{a}}\right)} e^{3 a w_{a}}\right], 
\end{align}
    where we have also included the Chevallier-Polarski-Linder (CPL) parameterization~\citep{2001IJMPD..10..213C, 2003PhRvL..90i1301L} of dynamical dark energy component whose equation of state is $w(z)=w_{0}+w_{a} z/(1+z)$~\citep{2008PhRvD..78b3526L}.}. When we increase the value of $\Omega_\nu$ we reduce the value of $\Omega_{\mathrm{cdm}}$ by the same amount in order to maintain a  flat cosmology. When we vary $M_\nu$ in our simulations we kept fixed the value of the power spectrum initial amplitude, $A_s$, therefore varying $M_\nu$ will result in different values for $\sigma_{8}$ at $z=0$.\\
    
    \item $h$: The dimensionless Hubble-Lemaitre parameter, which sets the value of the Hubble-Lemaitre constant, i.e. $H_{0}=100\, h$~$\mathrm{km}$~$\mathrm{s}^{-1}$~$\mathrm{Mpc}^{-1}´$ at $z=0$.\\
    
    \item $\Omega_{\mathrm{b}}$: The baryon density parameter, $\Omega_{\mathrm{b}}\equiv\rho_{\mathrm{b}}/\rho_\mathrm{c}$. Changes to $\Omega_{\mathrm{b}}$ are compensated by changing $\Omega_{\mathrm{cdm}}$ such that $\Omega_{\textrm{m}}$ remains constant.\\
\end{enumerate}

\begin{table}
\begin{center}
\caption{Our four ``reference'' simulations (\textit{Nenya}, \textit{Narya}, \textit{Vilya} and \textit{The One}) share the following cosmological parameters: $\sigma_{8} = 0.9$, $M_{\nu} = 0$, $w_{0} = -1$ and $w_\textrm{a} = 0$. The parameters listed below have been varied.}
\begin{tabular}{|c|c|c|c|c|c|}
\hline
\textbf{Name} & $\Omega_\textrm{m}$ & $n_\textrm{s}$ & $h$ & $\Omega_\textrm{b}$ & $m_{\rm DM}$ $[h^{-1} {\rm M_\odot}]$ \\ \hline
\textbf{\textit{Nenya}}           & 0.315               & 1.01           & 0.60         & 0.050               & $10^{9.51}$ \\ 
\textbf{\textit{Narya}}           & 0.360               & 1.01           & 0.70         & 0.050               & $10^{9.57}$ \\ 
\textbf{\textit{Vilya}}           & 0.270               & 0.92           & 0.65         & 0.060               & $10^{9.44}$ \\ 
\textbf{\textit{The One}}           & 0.307               & 0.96           & 0.68         & 0.048               & $10^{9.5}$ \\ \hline
\end{tabular}
\label{tab:reference}
\end{center}
\end{table}

\begin{table*}
\caption{The values of the cosmological parameters that are modified for each simulation. All runs have the same cosmological parameters as those used for one of the four reference simulations listed in  \Tab{tab:reference} but with one parameter modified to match the values listed below. For example, the run referred to in the upper-left entry adopts cosmological parameters consistent with the {\em Nenya} simulation, but with a lower value of the rms density fluctuation amplitude, i.e. $\sigma_8=0.730$.}
\begin{tabular}{|c|c|c|c|c|c|c|c|}
\hline
Ref - \textbf{$\sigma_8$} & Ref - \textbf{$\Omega_\textrm{m}$} & Ref - \textbf{$n_\textrm{s}$}  & Ref - \textbf{$w_0$}  & Ref - \textbf{$w_\textrm{a}$}  & Ref - \textbf{$M_\nu$} & Ref - \textbf{$h$}   & Ref - \textbf{$\Omega_b$} \\ \hline
 \textit{Nenya} $0.730$     & \textit{Nenya} $0.23$      & \textit{The One} $0.920$ & \textit{Nenya} $-0.70$ & \textit{Nenya} $-0.30$ & \textit{Nenya} $0.1~\textrm{eV}$    & \textit{Nenya} $0.65$ & \textit{Nenya} $0.040$     \\ 
 \textit{The One} $0.770$     & \textit{Nenya} $0.27$      & \textit{The One} $0.940$ & \textit{Nenya} $-0.85$ & \textit{Nenya} $-0.15$ & \textit{Nenya} $0.2~\textrm{eV}$    & \textit{Narya} $0.70$ & \textit{Nenya} $0.045$     \\ 
 \textit{Nenya} $0.815$     & \textit{Narya} $0.36$      & \textit{The One} $0.965$ & \textit{Nenya} $-1.15$ & \textit{Nenya} $0.15$ & \textit{Nenya} $0.3~\textrm{eV}$    & \textit{Narya} $0.75$ & \textit{Nenya} $0.055$     \\ 
 \textit{Nenya} $0.860$     & \textit{Narya} $0.40$      & \textit{Narya} $0.990$ & \textit{Nenya} $-1.30$ & \textit{Nenya} $0.30$ & \textit{Nenya} $0.4~\textrm{eV}$    & \textit{Narya} $0.80$ & \textit{Nenya} $0.060$     \\ \hline
\end{tabular}
\label{tab:simulations}
\end{table*}

Our suite of simulations is designed around four reference runs, which we refer to as \textit{Nenya}, \textit{Narya}, \textit{Vilya} and \textit{The One}. All reference simulations share a number parameters -- specifically, $\sigma_8=0.9$, $M_\nu = 0.0~\textrm{eV}$, $w_0=-1.0$, $w_a=0.0$, and $L_\textrm{box}=512~\textit{h}^{-1}\textrm{Mpc}$ are the same for all of them -- but other parameters are varied as described in \Tab{tab:reference}. Along with these reference runs, we carried out 32 additional simulations divided in 8 groups (with 4 simulations in each group) according to the cosmological parameter that was varied. For the runs in a given group we uniformly vary a particular cosmological parameter so that it spans a $5\sigma$ or $10\sigma$ region (depending on the parameter) around the best-fit parameter values provided by~\cite{2020A&A...641A...6P}. For the case of the Hubble-Lemaitre parameter, $h$, we explore values that span a $4\,\sigma$ region around the best-fit value obtained from low-redshift supernovae data~\cite{2016ApJ...826...56R}.The selection of these cosmologies in particular is motivated by the criteria set forth in \cite{2020MNRAS.499.4905C}. In each simulation we modified only one cosmological parameter and keep all others fixed with respect to the values used for one of the reference simulations. The various runs are listed in \Tab{tab:simulations}, where the column headers indicate the cosmological parameter that was modified, and the prefix indicates the reference model. 

All simulations were carried out using a lean version of \texttt{L-Gadget3} \citep[see][]{2008MNRAS.391.1685S, 10.1111/j.1365-2966.2012.21830.x} and evolved the DM density field using ${\rm N_{DM}}=1536^3$ equal-mass DM particles; they all employed the same softening length: $\epsilon=5\,h^{-1}\mathrm{kpc}$. All simulation volumes are approximately $V_{\textrm{box}}\approx(512\,h^{-1}\mathrm{Mpc})^3$, but vary slightly from run to run\footnote{Slight differences in the box size between the various runs ensures that variations of our reference models, when the cosmology-rescaling algorithm is employed to match the corresponding reference cosmology, will have a volume of exactly $V_{\textrm{box}}=(512\,h^{-1}\mathrm{Mpc})^3$.}. The slight variation in box size, along with changes to $\Omega_\mathrm{m}$, result in small differences in the DM particle masses between simulations. Our lowest-mass resolution run has
$m_{\rm DM}=10^{10.01}\,h^{-1}{\rm M_\odot}$ (\textit{Extreme high}-$n\textrm{s}$), and our highest mass-resolution run has $m_{\rm DM}=10^{9.41}\,h^{-1}{\rm M_\odot}$ (\textit{Extreme low}-$h$); the particle masses of all other simulations falls within this range. We use a version of NgenIC~\citep{2015ascl.soft02003S} that employs second-order Lagrangian Perturbation Theory (2LPT) to generate the initial conditions at $z=49$ for each simulation.

For simulations including massive neutrinos, we created initial conditions using the scale dependent backscaling technique described in \cite{2017MNRAS.466.3244Z}. We then evolved these simulations with the neutrino implementation of \cite{2013MNRAS.428.3375A}, where neutrino perturbations were solved on a grid, employing a linear response function that is sensitive to the non-linearities developed in the cold matter distribution.

To reduce cosmic variance, we followed the approach of \cite{10.1093/mnrasl/slw098} and carried out paired-phase counterparts of each of our simulations, which doubles the total number of simulations used in our analysis. We differentiate the two simulations within each of the fixed-paired doublets with the suffixes ``- $0$'' and ``- $\pi$''. For more information regarding the fixing and pairing technique and how it reduces cosmic variance in cosmological simulations see \cite{10.1093/mnrasl/slw098}, \cite{2019MNRAS.487...48C}, \cite{2021arXiv210313088K} and \cite{2022arXiv220403868M}.

We identify haloes and subhaloes in our simulations using a Friends-of-Friends algorithm~\citep{1985ApJ...292..371D}, with linking length $b=0.2$, and a modified version of \texttt{SUBFIND}~\citep{10.1046/j.1365-8711.2001.04912.x}. As discussed in~\cite{2020MNRAS.499.4905C}, our implementation of \texttt{SUBFIND} is able to robustly identify substructure haloes by considering their prior evolution.

We construct merger trees by linking haloes and subhaloes between consecutive snapshots, starting from the first snapshot in which a particular halo is identified. We then progress through subsequent snapshots and  determine which halo or subhalo is its most likely descendant. To do so, we track its 15 most-bound particles between snapshots and identify all (sub)haloes in which these particles end up; these constitute a set of possible descendants. We identify the most likely "true" descendant by considering which (sub)halo candidate has the highest score based on the number of particles it inherits weighted by their rank ordered binding energy with respect to the original (sub)halo. This approach constitutes a slight modification to the method used by~\cite{2012MNRAS.426.2046A} where only the inherited number of (most-bound) particles is considered but not their binding energies.

\subsection{Halo dynamical state and relaxedness}\label{subsec_relaxedness}

In this work we analyze the $c(M,z)$ relations of "relaxed" DM haloes. We discard unrelaxed haloes from our analysis because their spherically averaged density profiles are likely to deviate from spherical symmetry, and as such be ill fit by simple analytic profiles such as NFW or Einasto.

Following~\cite{2012MNRAS.427.1322L}, we consider a halo unrelaxed if its half-mass formation lookback time (since identification), i.e. $t_{\mathrm{h}} = t_{\mathrm{lb}}(z_\mathrm{h}) - t_{\mathrm{lb}}(z_0)$, is less than a crossing time, $t_\mathrm{cross}= 2\,r_{200} / V_{200}$. Following \cite{2007MNRAS.381.1450N}, we also discard haloes for which the distance between their center of mass and the position of the gravitational potential minimum is greater than $0.07\, r_{200}$ as well as those whose substructure mass fraction (i.e. the mass contained in subhaloes within $r_{200}$ of the host halo) exceeds $0.1\, M_{200}$.

\subsection{Analysis of halo density profiles}\label{subsec_obtain_rs}

Much of our analysis focuses on the median mass-concentration-redshift relations obtained from the best-fit density profiles of well-resolved haloes in our simulations, which we initially compute using logarithmically spaced mass bins of width $\Delta\log M_{200}=0.1$ that span the range $M_{200}\in(10^{13}, 10^{15.2})\; \textit{h}^{-1}\mathrm{M}_\odot$. Following previous works \cite[e.g.,][]{2008MNRAS.387..536G, 2014MNRAS.441.3359D, 2018ApJ...859...55C, 2019MNRAS.488.3663L, 2020MNRAS.495.4994B}, we then discard bins corresponding to haloes with fewer $5000$ particles within their virial radius, $r_{200}$, as well as those containing fewer than $50$ haloes, the latter to avoid excessive noise in the relations. 

To compute the concentrations of haloes we fit each of their spherically-averaged density profiles to Einasto's formula, i.e. \Eq{eq_Einasto}, but fix the value of $\alpha$ according to the $\alpha-\nu$ relation obtained by ~\cite{2008MNRAS.387..536G}, i.e. 

\begin{equation}
    \label{eq_Gao}
    \alpha = 0.155 + 0.0095\,{\nu\left(M,z\right)}^2.
\end{equation}
When fitting the density profiles we discard radial bins that are below the resolution limit, $r_{\textrm{min}}$. We follow \citet{2003MNRAS.338...14P} and define $r_{\textrm{min}}$ as the radius at which relaxation time is equal to the circular orbital time at the virial radius, i.e. $t_{\textrm{relax}}(r_\textrm{min})=t_{\textrm{circ}}(r_{200})$ \citep[see also][]{2019MNRAS.487.1227Z, 2019MNRAS.488.3663L}. This yields the following condition:
\begin{equation}
    \frac{t_{\text {relax }}\left(r_{\text {min }}\right)}{t_{\text {circ }}\left(r_{200}\right)}=\frac{\sqrt{200}}{8} \frac{N\left(r_{\text {min }}\right)}{\ln N\left(r_{\text {min }}\right)}\left[\frac{\rho_{c}(z_0)}{{\rho_{\text{enc}}}\left(r_{\text {min }}\right)}\right]^{\frac{1}{2}}=1,
\end{equation}
where $\rho_{\text {c}}(z_0)$ is the critical density of the universe at the halo identification redshift $z_0$, and $N(r_{\text {min }})$ and ${\rho_\text{enc}}(r_{\text {min }})$ are the enclosed number of particles and enclosed density at $r_{\text{min}}$, respectively. We also discard radial bins for which $r>r_{\rm max}=0.8 r_{200}$, where density profiles can be sensitive to local departures from equilibrium \citep[see, e.g.][]{2020MNRAS.493.2926L}. When carrying out our fits, we restrict the best-fit value of $r_{-2}$ to the range $r_{\rm min}\leq r_{-2}\leq r_{\rm max}$.

Although we have excluded unrelaxed haloes from our analysis, we nonetheless encounter a great diversity in profile shapes, and for a number of them the best-fit value of $r_{-2}$ is equal to $r_\textrm{min}$ or $r_\textrm{max}$. For these cases, the true value of $r_{-2}$ is likely outside the resolved radial range and our estimate of $r_{-2}$ therefore represents a lower or upper limit. We surmount this problem by discarding all mass bins in which more than 30 per cent of haloes have either $r_{-2}=r_{\rm min}$ or $r_{-2}=r_{\rm max}$, which ensures that such poorly-fit systems do not bias the median concentrations used in our analysis. We have employed a simulation with higher resolution (more than $3$ times the number of particles and $50$ per cent smaller force softening) to verify that this procedure yields robust values for the median concentrations.

\begin{figure}
\begin{center}
\includegraphics[width=1.\columnwidth]{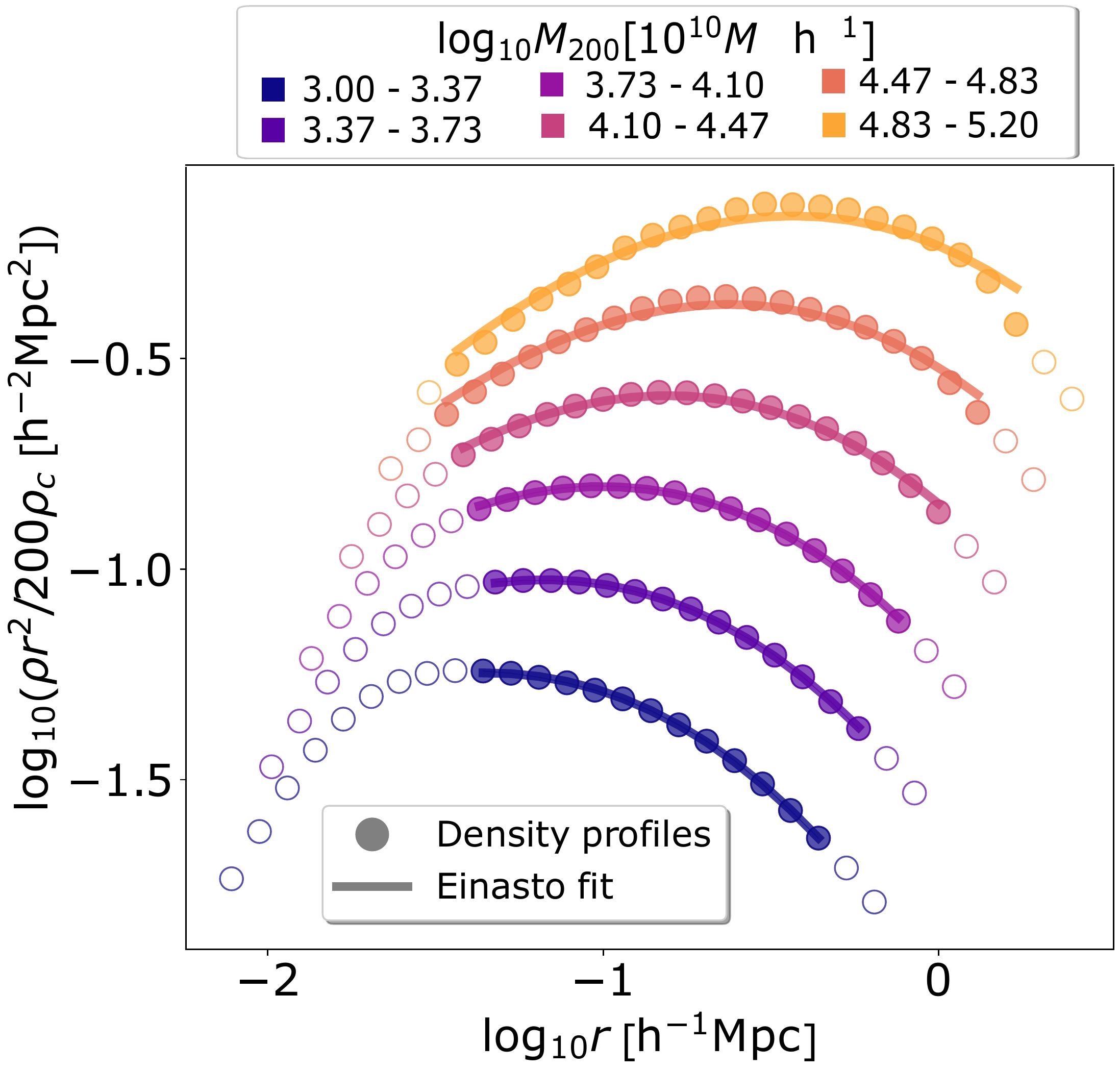}\vspace*{-0.2cm}
\caption{Median density profiles corresponding to six different logarithmically-spaced mass bins spanning the range $M_{200}\in(10^{13}, 10^{15.2})\; \textit{h}^{-1}\mathrm{M}_\odot$. All haloes were identified in the $\textit{The One}-\pi$ simulation at $z_0=0$ (circles). The filled circles correspond to the "resolved" radii used when carrying out our fits, i.e. they correspond to radial bins satisfying $r_{\textrm{min}} \leq r\leq r_{\textrm{max}}$ (see subsection~\ref{subsec_obtain_rs} for details). The thick solid lines show to the best-fit Einasto profiles with the values for $\alpha$ computed using \Eq{eq_Gao}. Different colors distinguish the different median virial masses, $M_{200}$, which are indicated in the legend in units of $\log_{10}M_{200}[10^{10} M_\odot \mathrm{h}^{-1}]$.}
\label{fig_density_profiles}
\end{center}
\end{figure}

In \Fig{fig_density_profiles} we show the median $z_0=0$ density profiles (weighted by a factor of $r^2$) for haloes of different virial mass in the $\textit{The One}-\pi$ simulation. Halo masses are logarithmically-spaced and span the range $M_{200}\in(10^{13}, 10^{15.2})\; \textit{h}^{-1}\mathrm{M}_\odot$. The filled circles correspond to radial bins with $r_{\rm min}\leq r\leq r_{\rm max}$. By plotting $\log_{10}(\rho r^2)$, the value of $r_{-2}$ is readily apparent as the radius of the "peak" of each best-fit profile. In addition to the median density profiles, we present their best Einasto fits (with $\alpha$ computed using \Eq{eq_Gao}; solid lines).

\begin{figure*}
\begin{center}
\includegraphics[width=\textwidth]{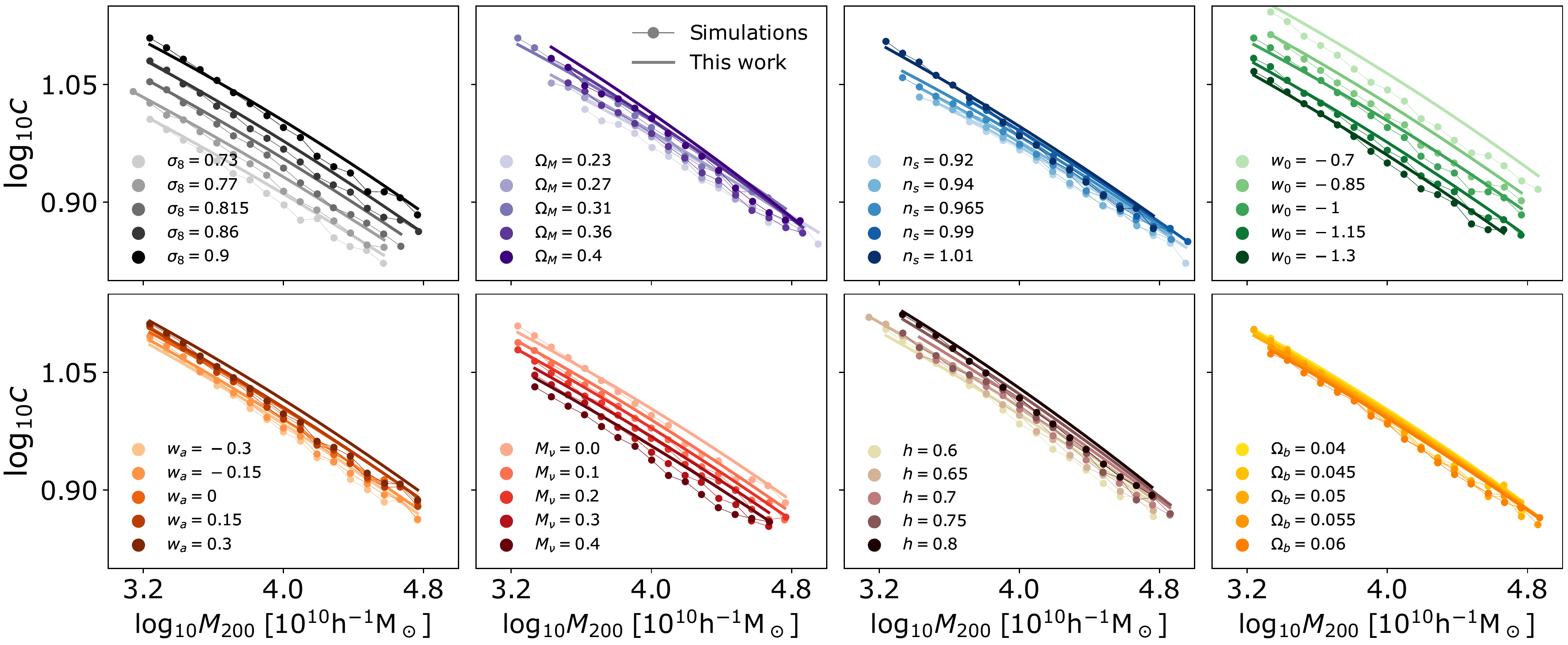}\vspace*{-0.2cm}
\caption{Median concentration-mass relations at $z_0=0$ for all cosmologies studied in this paper (see \Tab{tab:simulations}). Simulation results are shown as connected colored circles; the solid lines show the relations that are predicted by the model presented in this work, a modified version of the L16 model (using $A=493$; see Section~\ref{sec3} for details). From top-to-bottom and left-to-right, the cosmological parameters varied in each panel are, respectively, $\sigma_8$, $\Omega_\mathrm{m}$, $n_\mathrm{s}$, $w_0$, $w_\mathrm{a}$, $M_\nu, h$ and $\Omega_\mathrm{b}$. The simulation results correspond to the average of the median concentrations obtained for phase-$0$ and phase-$\pi$ simulations.}
\label{figc_M_relation}
\end{center}
\end{figure*}

%% file: sec3.tex
\section{Results}\label{sec3}

\subsection{Cosmology dependence of the mass-concentration-redshift relation}\label{subsec_c_M_relation}

In \Fig{figc_M_relation} we plot using connected circles the $c(M)$ relations obtained from our suite simulations at $z_0=0$. The results are split into different panels according to the cosmological parameter that was varied. With a total of 72 simulations, \Fig{figc_M_relation} represents, to our knowledge, the most extensive analysis to date of the cosmology-dependence of the mass-concentration relation. For each cosmology, we plot the average concentration of haloes in each mass bin after combining the fixed amplitude and inverted-phase simulations (all the results presented henceforth correspond to averages of our fixed-amplitude and inverted-phase simulations). For completeness, in Appendix~\ref{A3} we present the concentration-mass relations at $z_0=0.5$.

In agreement with previous findings, \Fig{figc_M_relation} shows that the concentrations of relaxed DM haloes decreases as a function of halo mass for all cosmological models studied. This is consistent with interpretation that structure forms hierarchically, i.e. low-mass haloes typically form before more massive ones, and that the concentrations of haloes are correlated with their formation times.

The results also illustrate how varying different cosmological parameters affects the concentration-mass relation. For example, regardless of halo mass, increasing the value of $\sigma_8$ leads to higher concentration. This is because higher $\sigma_8$ implies higher linear fluctuation amplitudes at fixed mass, and so earlier average formation times.

Higher values of $w_0$ also increase concentrations at all masses. This is because $w_0$ alters the growth histories of haloes through the dark energy term in~\Eq{eq_friedmann}. Specifically, higher $w_0$ leads to earlier halo formation times since (for a fixed value of $\sigma_8$) the increased contribution of dark energy to the universal expansion history demands that the haloes of a given mass form earlier, which in turn increases their concentration.

As a final example, consider the impact of $\Omega_{\rm b}$. For the runs plotted in the lower-right panel of \Fig{figc_M_relation}, $\Omega_{\rm b}$ contributes at least 4 per cent and at most 6 per cent of the critical density of the universe. Such a small contribution from baryons implies that the matter component in all our runs is dominated by cold dark matter. As such, the formation histories--and as a consequence, the concentrations--of haloes are largely insensitive to $\Omega_{\rm b}$, at least over the range of values studied here.

\subsection{The relationship between the characteristic densities of haloes and their formation histories}\label{subsec_rho_enc_rho_coll}

\begin{figure}
\begin{center}
\includegraphics[width=1.\columnwidth]{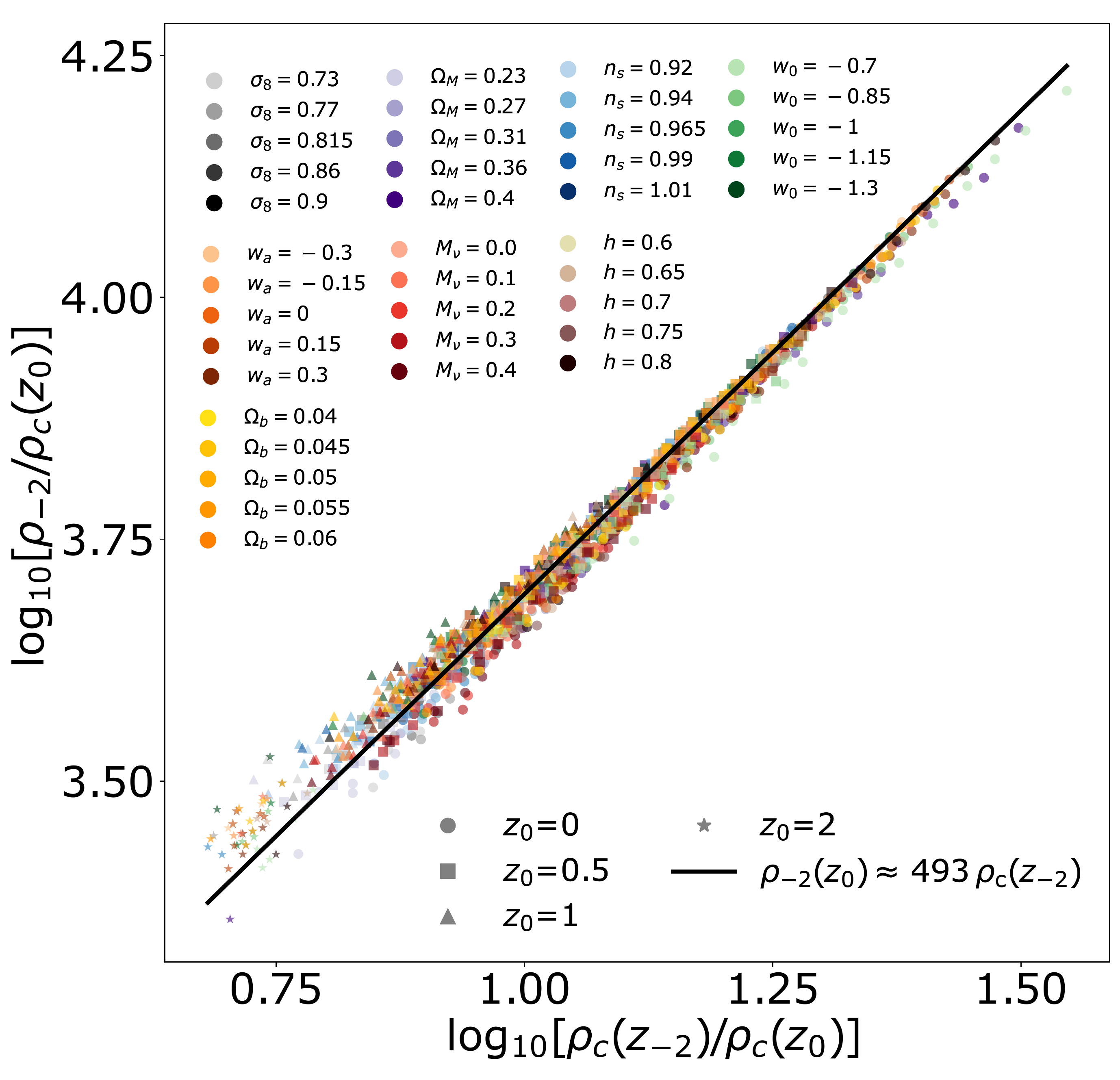}\vspace*{-0.2cm}
\caption{Relation between the median values of $\langle\rho_{\mathrm{-2}}\rangle$ and $\rho_{c}(z_\mathrm{-2})$ computed for relaxed haloes identified $z_0=0$, 0.5, 1 and 2 in our suite of simulations (the points correspond to median values obtained for equally-spaced logarithmic mass bins). The color of the points indicate the cosmological model and match the colors used for Fig.~\ref{figc_M_relation}. The shapes of the points indicate the redshift: $z_0=0$ as circles, $z_0=0.5$ as squares, $z_0=1$ as triangles, and $z_0=2$ as stars. The solid black line is the best fit to all the points: $\langle\rho_{\mathrm{-2}}\rangle  \approx 493\, \rho_\mathrm{c}(z_{-2})$.}
\label{fig_rho_enc_rho_coll_CMHs}
\end{center}
\end{figure}

As pointed out in \Sec{sec1}, there are a number models that aim to accurately predict the $c(M,z)$ relation, as well as its dependence on cosmological parameters. Many are based on empirical fits to results obtained from large suites of simulations \citep[e.g.][]{2014MNRAS.441.3359D, 2019ApJ...871..168D}, while others are based on physical models that relate the concentrations of haloes to their collapse histories \citep[e.g.][]{NFW, 1997ApJ...490..493N, 2001MNRAS.321..559B, 2002ApJ...568...52W, 2008MNRAS.387..536G, 2013MNRAS.432.1103L, 2014MNRAS.441..378L, 2015MNRAS.452.1217C, 2016MNRAS.460.1214L}. One model in particular, that of \citet[][L16 hereafter]{2016MNRAS.460.1214L}, has been shown to reproduce the mass-concentration relation for a variety of cosmological models, including cold and warm dark matter models that adopt sharply truncated power spectra \citep{2016MNRAS.460.1214L,2020Natur.585...39W,2022MNRAS.511.6019R}. The L16 model is based on the assumption (see appendix~\ref{A1_conc_model_theory} for more details) that the enclosed density within a halo scale radius, $\langle\rho_{-2}\rangle\equiv \langle\rho(r_{-2})\rangle$, is directly proportional to the critical density of the universe at the time when its characteristic mass, i.e. $M_{-2}\equiv M(<r_{-2})$, had first assembled into progenitors more massive than $0.02\times M_0$, where $M_0$ is the present-day mass of the halo. The redshift evolution of the mass fraction collapsed in such progenitors (i.e. those with masses exceeding $0.02\times M_0$) defines the halo's "collapsed mass history" (hereafter CMH for short). Below we test whether this result also holds for the various cosmologies explored in our simulation suite.

To do so, we use the simulated profiles to determine the mass $M_{-2}$ enclosed by the best-fit scale radius $r_{-2}$, and then define $\langle\rho_{-2}\rangle=3M_{-2}/4\pi r_{-2}^3$. The formation time, $z_{-2}$, is defined as the redshift at which the halo's CMH first exceeds \Menc, which is obtained by interpolating along the CMH that we calculated using each halo's merger tree. 

In \Fig{fig_rho_enc_rho_coll_CMHs} we plot the relation between $\langle\rho_{\mathrm{-2}}\rangle$ and $\rho_{c}(z_\mathrm{-2})$ for all the simulations described in \Sec{subsec_simulations}, and for redshifts $z_0=0,0.5,1,2$ (distinguished using different symbols). Each point corresponds to the average $\langle\rho_{\mathrm{-2}}\rangle$ and $\rho_{c}(z_\mathrm{-2})$ calculated for the same mass bins used to construct \Fig{figc_M_relation}. \Fig{fig_rho_enc_rho_coll_CMHs} reveals an approximate power-law relation between $\langle\rho_{\mathrm{-2}}\rangle$ and $\rho_{c}(z_\mathrm{-2})$ that is largely independent of cosmology, halo mass and redshift. Note too that the relation plotted has a "natural" slope very close to 1, i.e. $\langle\rho_{\mathrm{-2}}\rangle \propto \rho_{c}(z_\mathrm{-2})$. The solid line shows the best-fit relation: $\langle\rho_{\mathrm{-2}}\rangle \approx 493\, \rho_\mathrm{c}(z_{-2})$.

The existence of a tight relation between $\langle\rho_{\mathrm{-2}}\rangle$ and $\rho_{c}(z_\mathrm{-2})$ suggests that the concentrations of haloes -- regardless of mass, redshift, or cosmology -- can be predicted if an accurate model for the CMHs of haloes can be found. We investigate this next.

\begin{figure}
\begin{center}
\includegraphics[width=1.\columnwidth]{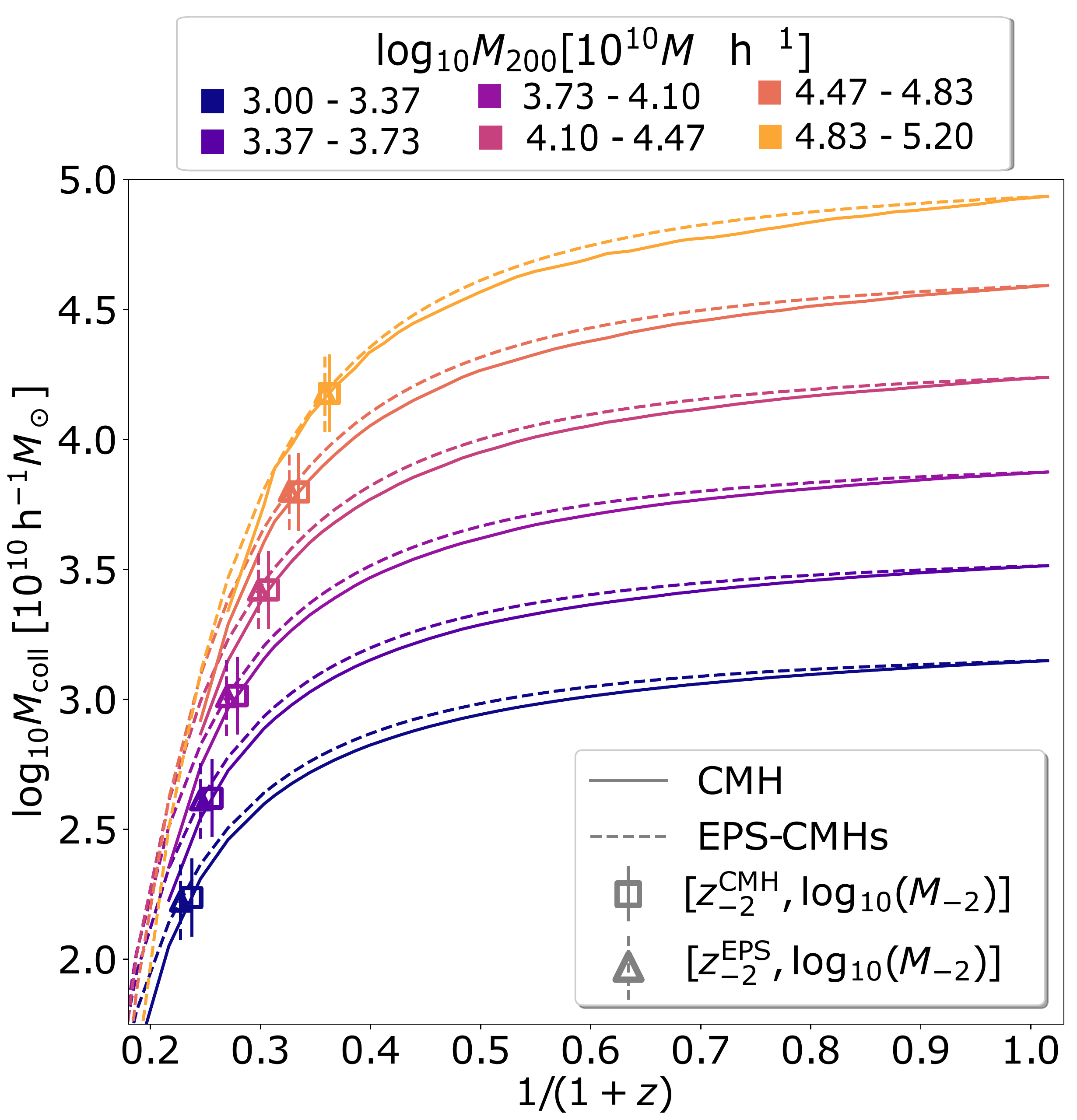}\vspace*{-0.2cm}
\caption{Median collapsed mass histories (i.e. $M_{\rm coll}$) for DM haloes with five different masses identified in the \textit{The One} simulation at redshift $z_0=0$ (solid lines; note that these are the same haloes whose density profiles are plotted in Fig.~\ref{fig_density_profiles}). Results are plotted as a function of scale factor, $a$. The formation time (defined as the point at which $M_{\rm coll} = M_{-2}$) for each halo is marked with a square crossed by a solid vertical segment (used for clarity) on top of the CMH. The dashed colored lines correspond to the CMHs predicted by the extended Press-Schechter theory (EPS-CMHs) and have been computed using \Eq{eq_EPS} with $\delta_{sc}=1.46$ and $f=0.02$. The open triangles crossed by dashed vertical segments indicate the formation times obtained from the EPS-CMHs.}
\label{figMCH_vs_EPS}
\end{center}
\end{figure}

\subsection{Predicted formation times based on the extended Press-Schechter formalism}\label{subsec_EPS_CMHs}

\begin{figure}
\begin{center}
\includegraphics[width=1.\columnwidth]{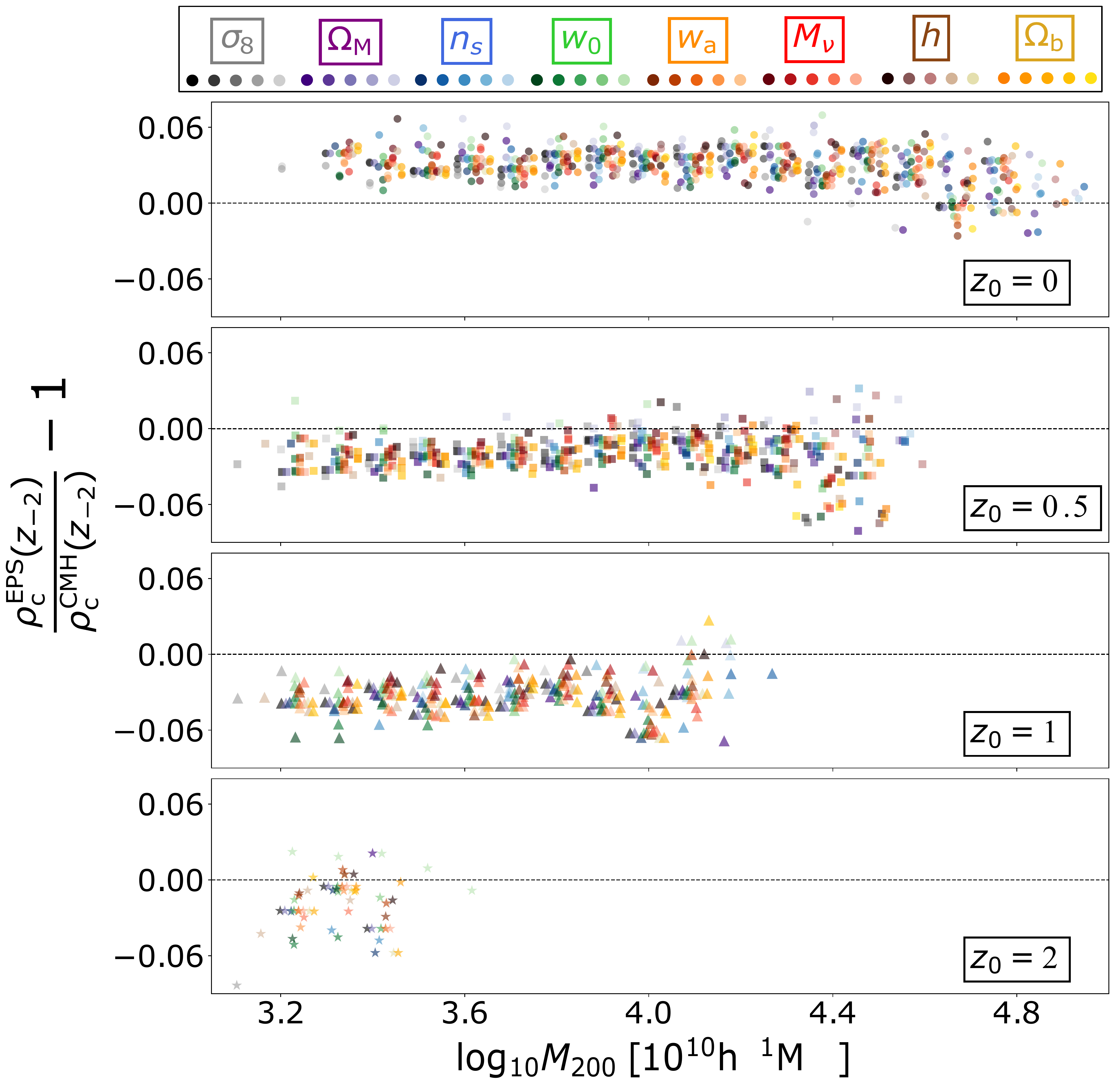}\vspace*{-0.2cm}
\caption{Relative difference between the median values of $\rho_{c}(z_\mathrm{-2})$ predicted using the EPS formalism, and measured from the simulated CMHs, plotted as a function of $M_{200}$. The results are split into four panels corresponding to redshifts $z_{0}= 0,0.5,1,$ and $2$. We present the results for all available cosmologies in our suite of simulations and color-code the points accordingly to the simulation they belong to following the color scheme adopted for \Fig{figc_M_relation}. The values associated with different
cosmologies are systematically shifted with respect to the mass-bin-centers
for better visualization.}
\label{fig_rho_coll_res}
\end{center}
\end{figure}

In \Fig{figMCH_vs_EPS} we show the median CMHs of haloes of different mass identified at $z_0=0$ in the $\textit{The One}-\pi$ simulation (solid lines; note that these are the same mass bins used to construct the density profiles plotted in \Fig{fig_density_profiles}). The outsized squares indicate the average halo formation times, $z_{-2}$, for the different mass bins. The dashed curves show, for comparison, the CMHs predicted by the extended Press-Schechter (EPS) formalism~\citep{1991ApJ...379..440B, 1993MNRAS.262..627L} for haloes of the same present day mass, see \Eq{eq_EPS}. The open triangles show the values of $z_{-2}$ associated with these EPS-collapsed mass histories (the latter referred to henceforth as EPS-CMHs). Note that the measured and predicted formation times agree quite well, as do the overall shapes of the CMHs. 

\begin{figure*}
\begin{center}
\includegraphics[width=\textwidth]{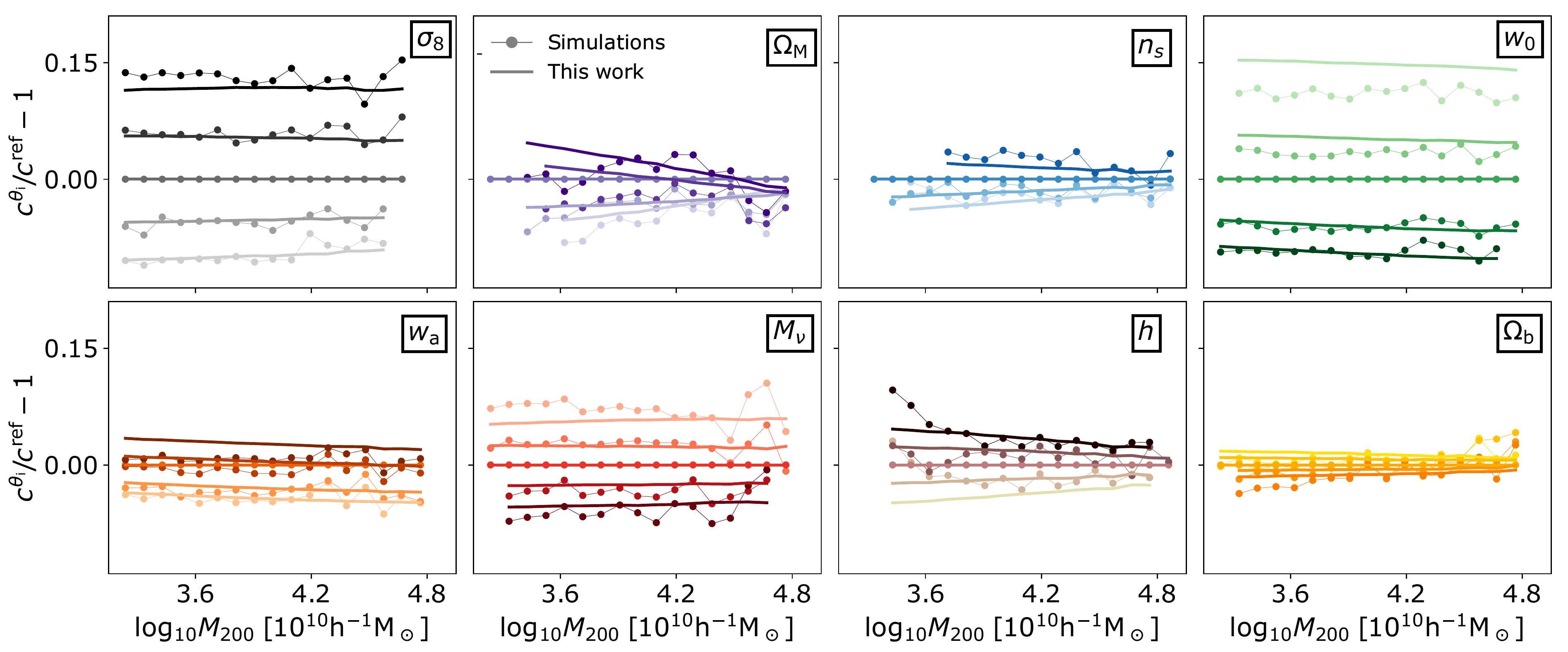}\vspace*{-0.2cm}
\caption{Relative differences between the $c(M,z)$ relation measured for different simulations (connected circles). The simulation taken as reference to compute the relative differences in each panel is the one with the intermediate value of the cosmological parameter that is varied. The plotting conventions match those used for \Fig{figc_M_relation}. The solid lines correspond to the relative differences between the predictions for the concentration computed using the re-calibrated L16 model.}
\label{figres_c_M_relation}
\end{center}
\end{figure*}

In \Fig{fig_rho_coll_res} we test how accurately the EPS model describes the formation times of haloes in our simulations (after a suitable modification to account for the impact of massive neutrinos in EPS, see Appendix~\ref{A1_conc_model_theory}). Here we plot the relative difference between the EPS-predicted formation redshifts, expressed as $\rho_\textrm{c}^{\textrm{EPS}}(z_{-2})$, and the formation redshifts measured directly from the simulated CMHs, i.e. $\rho_\textrm{c}^{\textrm{CMH}}(z_{-2})$. Each point corresponds to the median values of these quantities in bins of halo mass, and are plotted at different redshifts, which increase from the top to bottom panels. To help with visualization, we have applied a small horizontal shift to the values in each mass bin so that results obtained for different cosmological parameters can be easily distinguished. Our results show that the EPS-predicted formation redshifts agree well with the simulated ones, with residuals that show no clear systematic dependency on cosmology or on mass. But the residuals do exhibit a slight redshift dependence, but it remains below about 6 per cent for all models, mass bins and redshifts analyzed. Such small differences between the predicted and measured formation times of haloes do not significantly impact our ability to accurately model halo concentrations based on EPS CMHs, and we conclude that the CMHs of cold dark matter haloes can reliably modelled using the EPS formalism for a wide range of cosmological models.

\subsection{Model predictions for the mass-concentration-redshift relation}\label{subsec_c_vs_theta}

We follow L16 and use the power-law relation between $\langle\rho_{\mathrm{-2}}\rangle$ and $\rho_{c}(z_\mathrm{-2})$ presented in \Fig{fig_rho_enc_rho_coll_CMHs}, together with EPS-predicted formation times to predict the cosmology-dependence of the $c(M,z)$ relation. The results are plotted in \Fig{figc_M_relation} as solid colored lines, which agree well with the results of our simulations.

\Fig{figres_c_M_relation} further explores the extent to which the L16 model captures the correct cosmology- and mass-dependence of the $c(M,z)$ relation. The plot is organized to match Fig.~\ref{figc_M_relation}, with each panel showing results obtained from runs that vary a particular cosmological parameter; all results are plotted at $z_0=0$. The various connected circles show the relative differences between the median concentrations in each simulation measured with respect to those obtained from the run that was carried out with the intermediate value of the relevant cosmological parameter. The solid lines show the predictions of the L16 model, which reproduces the cosmology-dependence of concentration-mass relation rather well.

\begin{figure*}
\begin{center}
\includegraphics[width=\textwidth]{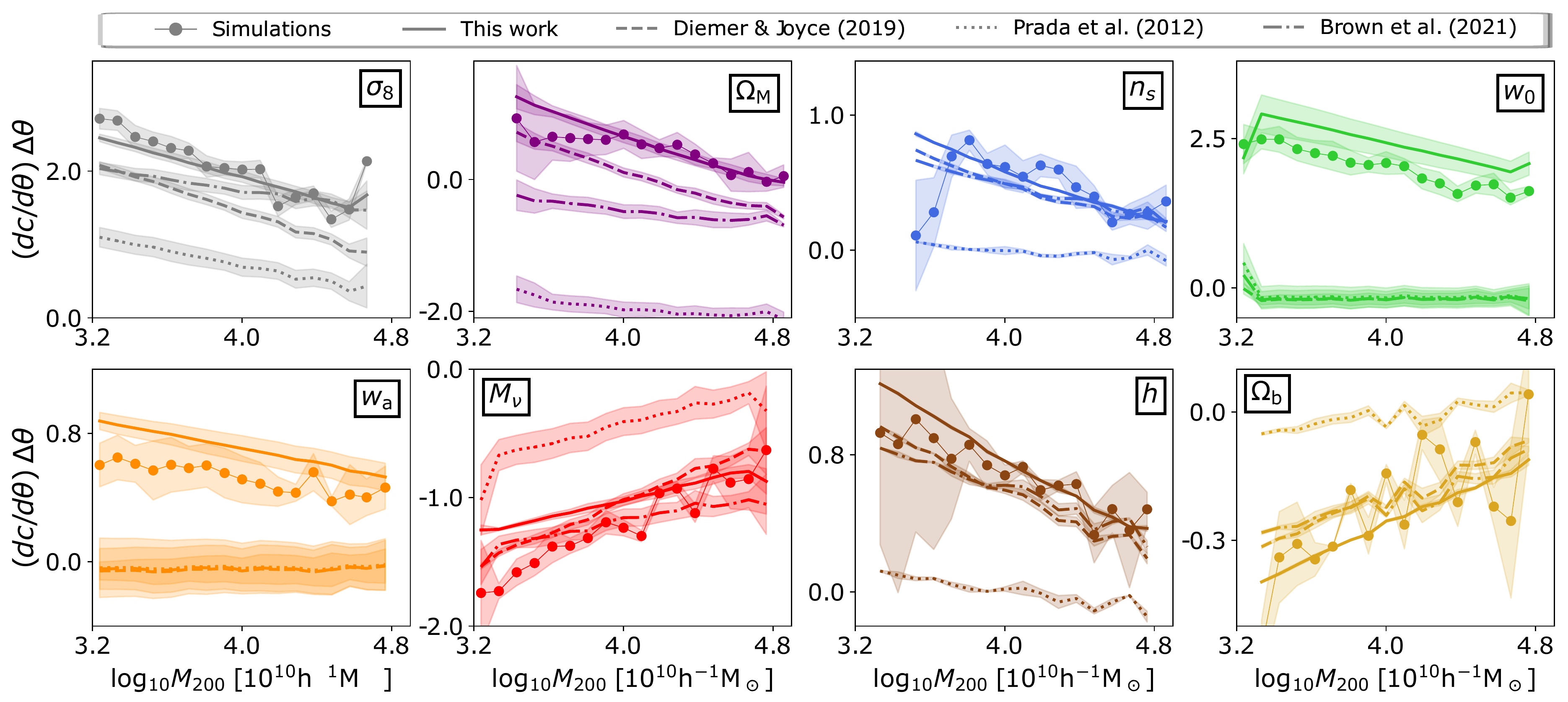}\vspace*{-0.2cm}
\caption{Linear dependence of the concentration on different cosmological parameters as a function of $M_{200}$. The results have been computed as described in subsection~\Sub{subsec_c_vs_theta}. The connected dots correspond to the results derived from our simulations. Different lines correspond to the results derived from different models: re-calibrated L16 model (solid lines), \citealt{2012MNRAS.423.3018P} (P12 -- dotted lines), \citealt{2019ApJ...871..168D} (DJ19 -- dashed lines), \citealt{2022MNRAS.509.5685B} (B22 -- dashed-dotted lines).}
\label{figdcdtheta_M_relation}
\end{center}
\end{figure*}

In \Fig{figdcdtheta_M_relation} we compare how well our measurements for the $c(M,z)$ relation can be reproduced by various other published concentration models. To produce \Fig{figdcdtheta_M_relation} we select the values for the concentration measured for each mass bin (considering separately the simulations in each subpanel of \Fig{figres_c_M_relation}), then, we obtain the gradient of the concentration with respect to the cosmological parameter that is varied, $dc / d\theta$, by fitting the selected points to a straight line. We repeat the process for all mass bins. The results obtained are then normalized by dividing by the interval spanned in each subpanel by the cosmological parameter that is been varied, $\Delta\theta$ (connected circles). We repeat this operation employing the predictions for the concentration provided by the re-calibrated L16 model (solid lines), the~\cite{2012MNRAS.423.3018P} model (``P12'', dotted lines), the~\cite{2019ApJ...871..168D} model (``DJ19'', dashed lines), and the the~\cite{2022MNRAS.509.5685B} model (``B22'', dashed-dotted lines).

The shaded regions around the trends in \Fig{figdcdtheta_M_relation} correspond to the error associated with the linear fit to obtain $dc / d\theta$. To a large extent the errors result from the scatter in the measured values for concentration.

The model that best captures the dependence of concentration on cosmology is our implementation of L16. Nevertheless is important to point out that the comparison of our results with the predictions provided by P12, DJ19 and B22 is somewhat unfair since, for instance, P12 aims to predict the concentrations for all halos (including unrelaxed ones). Regardless of this, it is important to note that the P12, DJ19 and B22 models do not predict any dependence of concentration on $w_0$ and $w_a$, whereas L16 provides reasonably accurate predictions. These results are not unexpected. The P12 and B22 models depend only on the shape of the (smoothed) density fluctuation power spectrum, but not on the assembly histories of haloes. Their predictions are therefore insensitive to the expansion history of the universe. The DJ19 model, however, does consider the slope of the growth factor (instead of the full merger history of haloes) when predicting halo concentrations, but this is largely insensitive (specially at low redshifts) to varying $w_0$ and $w_a$.

%% file: sec4.tex
\section{Application of the L16 model to scaling algorithms}\label{sec4}

In this section we illustrate how the L16 model can be used in studies that require a theoretical model capable of producing accurate concentration predictions. We will provide, as an example, the performance of the scaling algorithms (briefly summarized in the next paragraph), where the results substantially improve when including a concentration correction.

The scaling algorithm is a method developed by~\cite{10.1111/j.1365-2966.2010.16459.x} which allows one to rapidly generate mock or synthetic cosmological simulations from a "template" N-body simulation. The mock simulation that the algorithm generates contains the DM particles of the original simulation displaced to new positions in such a way that its density field accurately reproduces that of an actual N-body simulation executed using different cosmological parameters from those of the original N-body simulation. \cite{2019MNRAS.489.5938Z} extended the cosmology-rescaling technique to provide predictions when considering a hot component of arbitrary mass, such as neutrinos.

\cite{2020MNRAS.499.4905C} showed that very accurate predictions for the halo clustering can be achieved by including a concentration correction on top of the standard scaling algorithm. The concentration correction modifies the position of DM particles within haloes to match halo concentrations in the target cosmology.

\begin{figure}
\begin{center}
\includegraphics[width=1.\columnwidth]{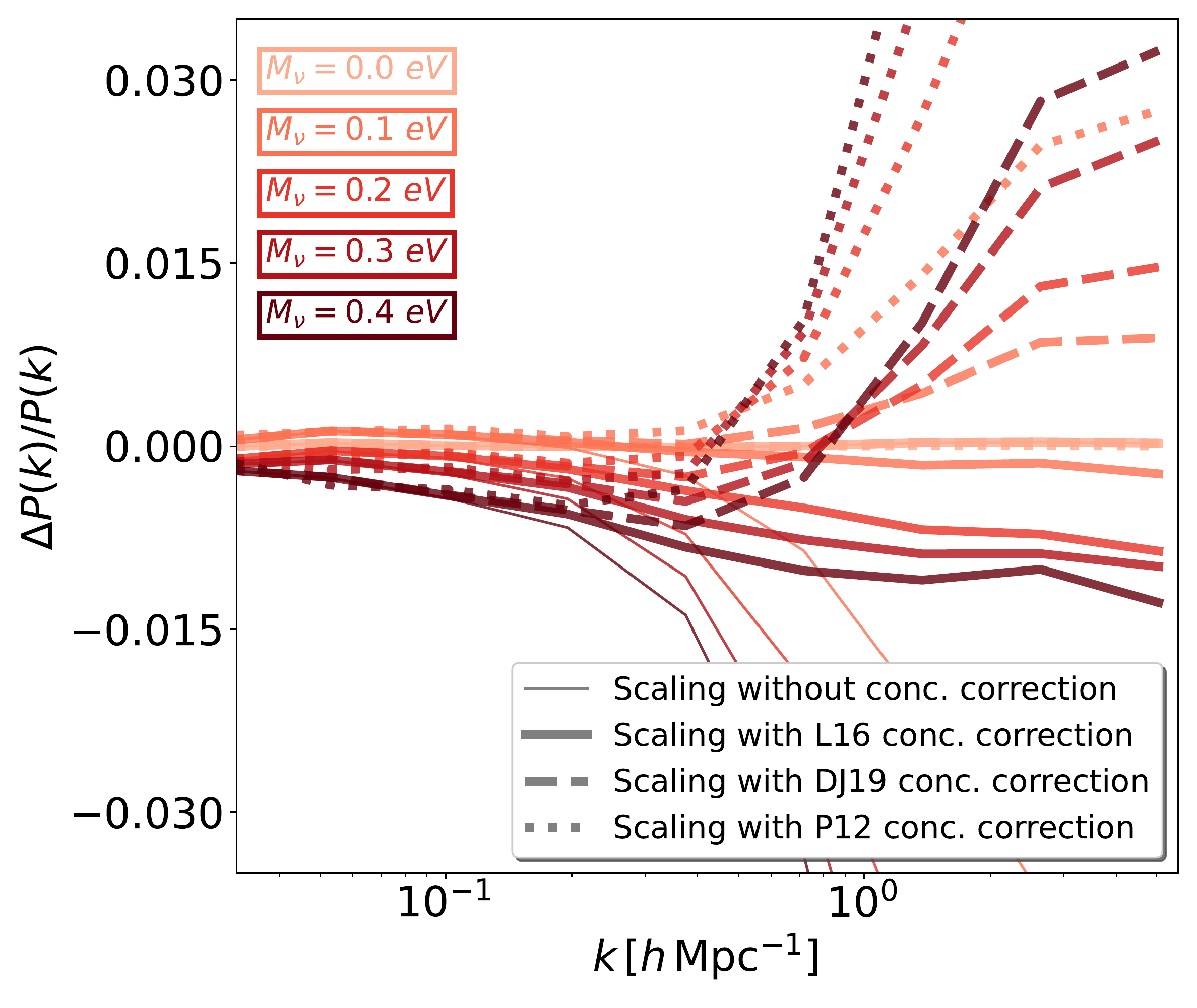}\vspace*{-0.2cm}
\caption{Relative difference between the power spectrum obtained from a gravity-only reference simulation and the power spectrum obtained from the corresponding rescaled simulation, i.e., $\Delta P(k)/P(k) =P_{\mathrm{scaled}}(k)/P_{\mathrm{N-body}}(k) -1$. We focus on models with different neutrino masses, and choose our reference simulation to be the one with $M_\nu=0\, eV$. The thin solid lines correspond to the case without applying any concentration correction and the other line styles correspond to concentration corrections obtained using three different concentration models: the re-calibrated L16 model (\Eq{eq_rho_enc_rho_coll}; thick solid lines), the \citealt{2012MNRAS.423.3018P} model (P12 -- dotted lines), and the model of \citealt{2019ApJ...871..168D} (DJ19 -- dashed lines). The different shades of red correspond to different neutrino masses. The Nyquist frequency corresponding to these set of simulations is $\log_{10}k_\mathrm{Ny} \, [h\,\mathrm{Mpc}^{-1}]\approx 0.97$, and the number of points selected to compute the power spectrum is sufficiently large so that aliasing effects are not important.}
\label{figrescaling}
\end{center}
\end{figure}

\Fig{figrescaling} illustrates how concentration corrections improve the accuracy of the power spectrum corresponding to a scaled simulation generated using the scaling algorithm. To generate this figure we employ the set of simulations presented in ~\S\ref{subsec_simulations} in which we vary the total neutrino mass, $M_\nu$, from $0.0\,\mathrm{eV}$ to $M_\nu = 0.4\,\mathrm{eV}$, keeping all other cosmological parameters (those of \textit{Nenya}) fixed, see \Tab{tab:simulations}.

We apply a scaling algorithm to the N-body simulation with $M_\nu = 0\, \mathrm{eV}$ to produce mock simulations that mimic the behaviour of runs with $M_\nu = 0.1, 0.2, 0.3, 0.4\, \mathrm{eV}$. We first scale the $M_\nu = 0\, \mathrm{eV}$-simulation to the target cosmologies (changing $M_\nu$ to $0.1, 0.2, 0.3, 0.4\, \mathrm{eV}$ subsequently) without considering concentration corrections, then, we repeat the process employing three different concentration models -- the re-calibrated L16 model, the model presented in \cite{2012MNRAS.423.3018P} and the one from \cite{2019ApJ...871..168D} -- to provide the predictions for the concentration corrections.

We compute the power spectrum for the original N-body simulations and the scaled simulations with and without concentration correction. The relative differences between the original and scaled power spectra $\Delta P(k)/P(k) =P_{\mathrm{scaled}}(k)/P_{\mathrm{N-body}}(k) -1$ are plotted in \Fig{figrescaling} as a function of scale. The thin solid lines correspond to the comparison with respect to mock simulations scaled without concentration corrections; the remaining lines correspond to the comparison with scaled simulations in which we have considered the concentration corrections associated with different concentration models: the re-calibrated L16 model (solid lines), \citealt{2012MNRAS.423.3018P} (P12 -- dotted lines) and \citealt{2019ApJ...871..168D} (DJ19 -- dashed lines).

In \Fig{figrescaling} one can appreciate that the power spectra of the scaled simulations with concentration corrections are closer to the power spectra of the original N-body simulations in comparison with the case without concentration corrections. Different concentration models produce different levels of concentration corrections in the scaling technique which can be observed at the power spectrum level. The re-calibrated L16 model (i.e. \Eq{eq_rho_enc_rho_coll}) yields the most accurate predictions for the power spectrum when compared to the other models. In the most extreme scenario, when $M_\nu = 0.4 \,\mathrm{eV}$, the relative difference between the power spectrum from the rescaled simulation (for the L16 model) and the original simulation at $k \approx 4\,h\,\mathrm{Mpc}^{-1}$ is less than $1.5\%$; for the other concentration models the relative differences at this scale are at least twice as large.

%% file: sec5.tex
\section{Conclusions}\label{sec5}

In this paper, we carried out an extensive analysis of the cosmology dependence of the mass-concentration-redshift relation, $c(M,z)$, for dynamically relaxed dark matter halos. Our results were based on a large suite of gravity-only simulations in which we systematically varied the following cosmological parameters: $\sigma_{8}$, $\Omega_{\mathrm{M}}$, $\Omega_{\mathrm{b}}$, $n_\mathrm{s}$, $h$, $M_\nu, w_0$ and $w_\mathrm{a}$. Each parameter was varied linearly across a range that spans a 5 to 10$\sigma$ region (depending on the parameter; see \Tab{tab:simulations} and \Tab{tab:reference}) surrounding the best-fit value obtained by \cite{2020A&A...641A...6P}.

In agreement with previous work, we find that, regardless of the cosmological parameter varied, the concentrations of DM haloes, on average, decrease with increasing halo mass at fixed redshift (\Fig{figc_M_relation}), as well as with increasing redshift at fixed halo mass (\Fig{figc_M_relation_z_05}). For the range of parameter values we considered, concentrations are most sensitive to changes in $\sigma_8$, the rms amplitude of linear density fluctuations; they are least sensitive to changes in $\Omega_{\rm b}$, the baryon density parameter. This result is not surprising given the strong dependence of halo formation times on $\sigma_8$ and their weak dependence on $\Omega_{\rm b}$.

In general, our results agree with previous studies showing that the structure of dark matter haloes is strongly correlated with their formation histories \citep[e.g.][]{2014MNRAS.441..378L,2016MNRAS.460.1214L,2022arXiv220504474L}. Specifically, we find that halo concentrations, when expressed in terms of the enclosed density within the halo scale radius, i.e. $\langle\rho_{\mathrm{-2}}\rangle$, correlate strongly with the critical density at their formation time $z_{-2}$, i.e. $\rho_{c}(z_\mathrm{-2})$. Indeed, when the latter is defined as the point at which the "collapsed mass history" (CMH; defined as the mass in collapsed progenitors larger than a fraction $f=0.02$ of the halo's present day mass) first exceeds the halo's characteristic mass, i.e. $M_{-2}=M(<r_{-2})$, we find an approximately linear relation between the two densities that may be accurately approximated by 
\begin{equation}
    \langle\rho_{-2}\rangle=493 \times \rho_c(z_{-2}).
\label{eq:rho}  
\end{equation}
This simple relation holds for all cosmologies, redshifts, and masses studied. This is a somewhat surprising result and the most important finding of our paper: The relation between nonlinear halo structure and formation time is universal hinting that it may be a fundamental consequence of gravitational dynamics and collapse. This universality implies that our predictions for the concentration-mass relation should be valid even for cosmologies and halo masses outside the range considered here. 

We showed that equation~\ref{eq:rho}, when combined with an accurate model for halo CMHs based on extended Press-Schechter theory (see \Fig{figMCH_vs_EPS} and appendix~\ref{A1_conc_model_theory}), can be used to make accurate prediction for the mass-, cosmology- and redshift-dependence of halo concentrations (\Fig{figres_c_M_relation}) even when considering dynamical dark energy and massive neutrinos. We compared our predictions for the $c(M,z)$ relation with other published models (\Fig{figdcdtheta_M_relation}) and verified that they more accurately capture its cosmology dependence.

Our results confirm and extend those originally obtained by~\cite{2016MNRAS.460.1214L} and suggest that equation~\ref{eq:rho} can be used to accurately predict the concentrations of DM halos in a wide range of scenarios. This can be very useful in many areas of cosmology, e.g., to improve cosmological rescaling algorithms \citep[see][\Fig{figrescaling} and Section~\ref{sec4}]{2020MNRAS.499.4905C}.

%% file: A1.tex
\section{description of L16 model}\label{A1_conc_model_theory}

Here we explain the main ideas behind the \cite{2016MNRAS.460.1214L} (L16) model for predicting the $c(M,z)$ relation. Throughout this work we carefully examined if the assumptions upon which L16 is founded are fulfilled or not for a variety of cosmologies, masses and redshifts. The original L16 paper shows that their model accurately predicts the $c(M,z)$ relation for relaxed haloes in different cosmologies, including both cold and warm dark matter scenarios; in this work we extend their analysis by considering a broader range of distinct cosmologies, including the effect of massive neutrinos and dynamical dark energy.

The L16 model is based on an empirical relation between $\rho_{-2}(z_0)$, i.e. the enclosed density of a halo within the scale radius, $r_{-2}$ measured at redshift $z_0$, and \rhocoll, i.e. the critical density of the universe defined at a suitable formation redshift, $z_{-2}$, The relation can be written as:

\begin{equation}\label{eq_rho_enc_rho_coll}
\rho_{-2}(z_0) = A \rho_{\textrm{c}}(z_{-2}),
\end{equation}
where A is a proportionality constant. In L16 the halo formation redshift, $z_{-2}$, is defined as the redshift at which the collapsed-mass history (CMH) of a halo\footnote{For a given halo identified at redshift $z_0$, the collapsed-mass history is defined as the sum of all the mass contained in progenitor haloes at redshift $z > z_0$ that end up being accreted by the halo of interest and whose mass exceeds a certain fraction $f$ (in L16 $f\equiv0.02$) of the halo's final mass. This can be calculated for simulated haloes using their merger trees, or predicted theoretically using the extended Press-Shcechter formalism using \Eq{eq_EPS}.} first exceeds $M_{-2}\equiv M(r<r_{-2})$, i.e., the mass enclosed within a sphere of radius $r_{-2}$, at the $z_0$, centered around the potential minimum of the halo analyzed.

If indeed \Eq{eq_rho_enc_rho_coll} is verified, we can predict the value of $\rho_{-2}(z_0)$  employing an analytical model capable of reproducing the CMH of a halo given its mass, which would allow us to infer \rhocoll. To obtain the synthetic CMHs, we make use of the extended Press-Schechter (EPS) formalism~\citep{1991ApJ...379..440B, 1993MNRAS.262..627L}, according to which the mass contained in progenitors more massive than a certain fraction $f$ of the final halo mass, $M_0$, at a given redshift, $z$, is given by:

\begin{equation}\label{eq_EPS}
    \mathrm{M}_{\mathrm{coll}}(z) = \mathrm{M}_0 \operatorname{erfc}\left\{\frac{\delta_{\mathrm{c}}\left(\frac{D(z,k_{f \mathrm{M}_0})}{D(z_0,k_{f \mathrm{M}_0})}-1\right)}{\sqrt{2\left[\sigma^{2}\left(k_{f\mathrm{M_0}},z_0\right)-\sigma^{2}\left(k_{\mathrm{M_0}},z_0\right)\right]}}\right\},
\end{equation}
where $\delta_{\mathrm{c}}$ is the linear threshold for non-linear collapse (it can be computed, for example, using the spherical top hat approximation); $\sigma^{2}\left(k_{\mathrm{M}},z\right)$ is the variance of the linear matter density field computed at redshift $z$ at the scale $k_{\mathrm{M}}$ associated with the mass $\mathrm{M}$ (assuming a sharp-k window function $W(x)=3(\sin x-x \cos x) / x^{3}$); and $D(z,k_{fM_{0}})$ is the growth factor computed at redshift $z$ for a scale $k_{fM_{0}}$ corresponding to the mass $fM_{0}$.

Note that in \Eq{eq_EPS} we evaluate the variance of the matter field and the growth factor at different scales. This is particularly important for calculating the CMHs in cosmologies with massive neutrinos, where the scale dependence of the growth factor can have a significant impact.


%% file: A2.tex
\section{$c(M)$ relation at $z_0 = 0.5$}\label{A3}

\begin{figure}
\begin{center}
\includegraphics[width=1.\columnwidth]{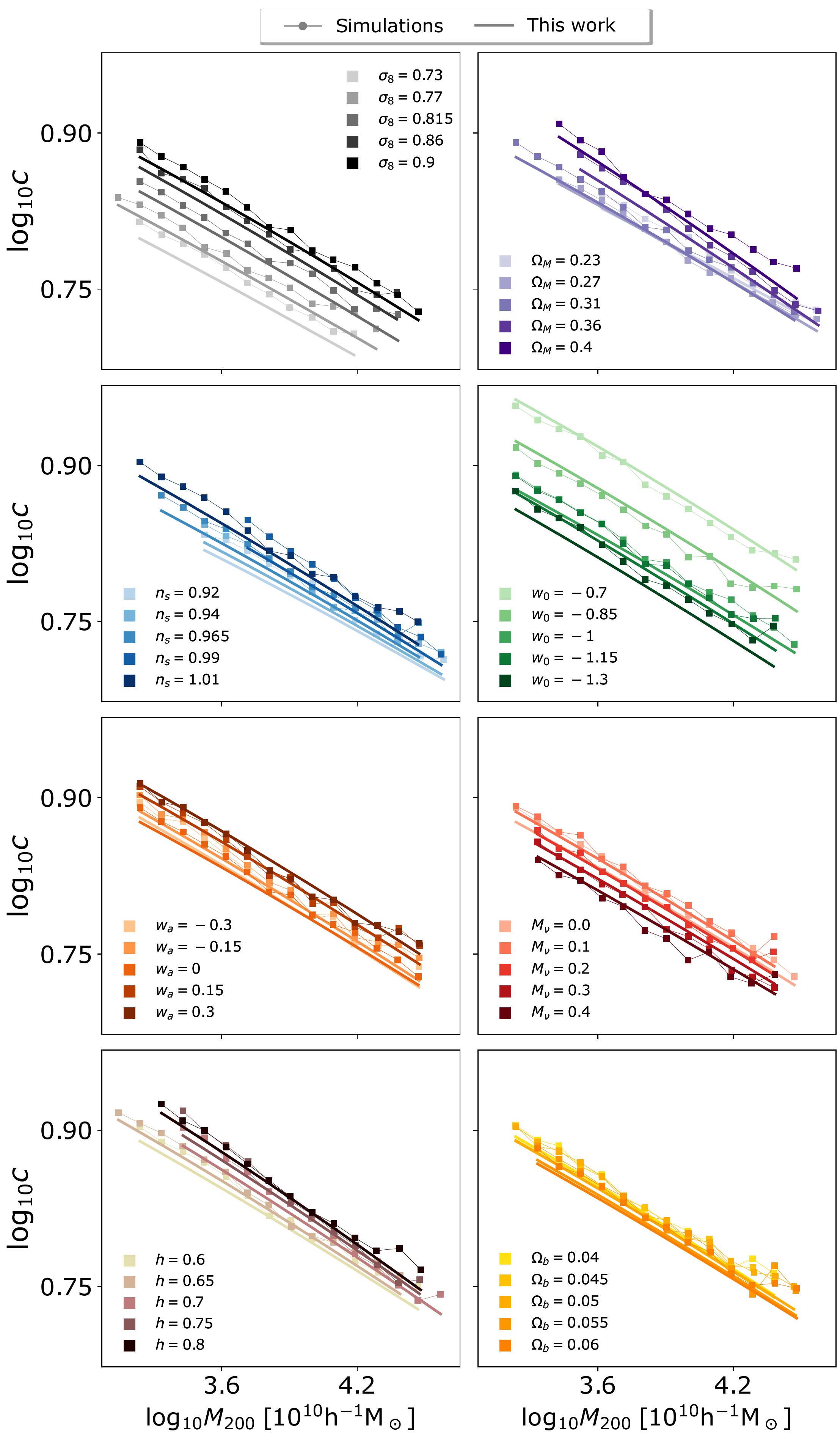}\vspace*{-0.2cm}
\caption{Concentration-mass relation for all cosmologies studied in this paper at redshift $z=0.5$ as a function of $M_{200}$ analogous to the results presented in \Fig{figc_M_relation}.}
\label{figc_M_relation_z_05}
\end{center}
\end{figure}

In \Fig{figc_M_relation_z_05} we present the results for the concentration-mass relation (as measured in \Fig{figc_M_relation}) at $z=0.5$ (connected squares). We also show the predictions provided by the L16 model at that redshift employing the same calibration as the one used in \Fig{figc_M_relation}.

%% file: main.bbl
\begin{thebibliography}{}
\makeatletter
\relax
\def\mn@urlcharsother{\let\do\@makeother \do\$\do\&\do\#\do\^\do\_\do\%\do\~}
\def\mn@doi{\begingroup\mn@urlcharsother \@ifnextchar [ {\mn@doi@}
  {\mn@doi@[]}}
\def\mn@doi@[#1]#2{\def\@tempa{#1}\ifx\@tempa\@empty \href
  {http://dx.doi.org/#2} {doi:#2}\else \href {http://dx.doi.org/#2} {#1}\fi
  \endgroup}
\def\mn@eprint#1#2{\mn@eprint@#1:#2::\@nil}
\def\mn@eprint@arXiv#1{\href {http://arxiv.org/abs/#1} {{\tt arXiv:#1}}}
\def\mn@eprint@dblp#1{\href {http://dblp.uni-trier.de/rec/bibtex/#1.xml}
  {dblp:#1}}
\def\mn@eprint@#1:#2:#3:#4\@nil{\def\@tempa {#1}\def\@tempb {#2}\def\@tempc
  {#3}\ifx \@tempc \@empty \let \@tempc \@tempb \let \@tempb \@tempa \fi \ifx
  \@tempb \@empty \def\@tempb {arXiv}\fi \@ifundefined
  {mn@eprint@\@tempb}{\@tempb:\@tempc}{\expandafter \expandafter \csname
  mn@eprint@\@tempb\endcsname \expandafter{\@tempc}}}

\bibitem[\protect\citeauthoryear{{Ali-Ha{\"\i}moud} \&
  {Bird}}{{Ali-Ha{\"\i}moud} \& {Bird}}{2013}]{2013MNRAS.428.3375A}
{Ali-Ha{\"\i}moud} Y.,  {Bird} S.,  2013, \mn@doi [\mnras]
  {10.1093/mnras/sts286}, \href
  {https://ui.adsabs.harvard.edu/abs/2013MNRAS.428.3375A} {428, 3375}

\bibitem[\protect\citeauthoryear{{Amorisco} et~al.,}{{Amorisco}
  et~al.}{2021}]{2021arXiv210900018A}
{Amorisco} N.~C.,  et~al., 2021, arXiv e-prints, \href
  {https://ui.adsabs.harvard.edu/abs/2021arXiv210900018A} {p. arXiv:2109.00018}

\bibitem[\protect\citeauthoryear{{Angulo} \& {Hahn}}{{Angulo} \&
  {Hahn}}{2022}]{AnguloHahn2022}
{Angulo} R.~E.,  {Hahn} O.,  2022, \mn@doi [Living Reviews in Computational
  Astrophysics] {10.1007/s41115-021-00013-z}, \href
  {https://ui.adsabs.harvard.edu/abs/2022LRCA....8....1A} {8, 1}

\bibitem[\protect\citeauthoryear{Angulo \& Pontzen}{Angulo \&
  Pontzen}{2016}]{10.1093/mnrasl/slw098}
Angulo R.~E.,  Pontzen A.,  2016, \mn@doi [Monthly Notices of the Royal
  Astronomical Society: Letters] {10.1093/mnrasl/slw098}, 462, L1

\bibitem[\protect\citeauthoryear{Angulo \& White}{Angulo \&
  White}{2010}]{10.1111/j.1365-2966.2010.16459.x}
Angulo R.~E.,  White S. D.~M.,  2010, \mn@doi [Monthly Notices of the Royal
  Astronomical Society] {10.1111/j.1365-2966.2010.16459.x}, 405, 143

\bibitem[\protect\citeauthoryear{Angulo, Springel, White, Jenkins, Baugh  \&
  Frenk}{Angulo et~al.}{2012a}]{10.1111/j.1365-2966.2012.21830.x}
Angulo R.~E.,  Springel V.,  White S. D.~M.,  Jenkins A.,  Baugh C.~M.,   Frenk
  C.~S.,  2012a, \mn@doi [Monthly Notices of the Royal Astronomical Society]
  {10.1111/j.1365-2966.2012.21830.x}, 426, 2046

\bibitem[\protect\citeauthoryear{{Angulo}, {Springel}, {White}, {Jenkins},
  {Baugh}  \& {Frenk}}{{Angulo} et~al.}{2012b}]{2012MNRAS.426.2046A}
{Angulo} R.~E.,  {Springel} V.,  {White} S.~D.~M.,  {Jenkins} A.,  {Baugh}
  C.~M.,   {Frenk} C.~S.,  2012b, \mn@doi [\mnras]
  {10.1111/j.1365-2966.2012.21830.x}, \href
  {https://ui.adsabs.harvard.edu/abs/2012MNRAS.426.2046A} {426, 2046}

\bibitem[\protect\citeauthoryear{{Bartelmann}, {Perrotta}  \&
  {Baccigalupi}}{{Bartelmann} et~al.}{2002}]{2002A&A...396...21B}
{Bartelmann} M.,  {Perrotta} F.,   {Baccigalupi} C.,  2002, \mn@doi [\aap]
  {10.1051/0004-6361:20021417}, \href
  {https://ui.adsabs.harvard.edu/abs/2002A&A...396...21B} {396, 21}

\bibitem[\protect\citeauthoryear{{Bode}, {Ostriker}  \& {Turok}}{{Bode}
  et~al.}{2001}]{2001ApJ...556...93B}
{Bode} P.,  {Ostriker} J.~P.,   {Turok} N.,  2001, \mn@doi [\apj]
  {10.1086/321541}, \href
  {https://ui.adsabs.harvard.edu/abs/2001ApJ...556...93B} {556, 93}

\bibitem[\protect\citeauthoryear{{Bond}, {Cole}, {Efstathiou}  \&
  {Kaiser}}{{Bond} et~al.}{1991}]{1991ApJ...379..440B}
{Bond} J.~R.,  {Cole} S.,  {Efstathiou} G.,   {Kaiser} N.,  1991, \mn@doi
  [\apj] {10.1086/170520}, \href
  {https://ui.adsabs.harvard.edu/abs/1991ApJ...379..440B} {379, 440}

\bibitem[\protect\citeauthoryear{{Brown}, {McCarthy}, {Diemer}, {Font},
  {Stafford}  \& {Pfeifer}}{{Brown} et~al.}{2020}]{2020MNRAS.495.4994B}
{Brown} S.~T.,  {McCarthy} I.~G.,  {Diemer} B.,  {Font} A.~S.,  {Stafford}
  S.~G.,   {Pfeifer} S.,  2020, \mn@doi [\mnras] {10.1093/mnras/staa1491},
  \href {https://ui.adsabs.harvard.edu/abs/2020MNRAS.495.4994B} {495, 4994}

\bibitem[\protect\citeauthoryear{{Brown}, {McCarthy}, {Stafford}  \&
  {Font}}{{Brown} et~al.}{2022}]{2022MNRAS.509.5685B}
{Brown} S.~T.,  {McCarthy} I.~G.,  {Stafford} S.~G.,   {Font} A.~S.,  2022,
  \mn@doi [\mnras] {10.1093/mnras/stab3394}, \href
  {https://ui.adsabs.harvard.edu/abs/2022MNRAS.509.5685B} {509, 5685}

\bibitem[\protect\citeauthoryear{{Bullock}, {Kolatt}, {Sigad}, {Somerville},
  {Kravtsov}, {Klypin}, {Primack}  \& {Dekel}}{{Bullock}
  et~al.}{2001}]{2001MNRAS.321..559B}
{Bullock} J.~S.,  {Kolatt} T.~S.,  {Sigad} Y.,  {Somerville} R.~S.,  {Kravtsov}
  A.~V.,  {Klypin} A.~A.,  {Primack} J.~R.,   {Dekel} A.,  2001, \mn@doi
  [\mnras] {10.1046/j.1365-8711.2001.04068.x}, \href
  {https://ui.adsabs.harvard.edu/abs/2001MNRAS.321..559B} {321, 559}

\bibitem[\protect\citeauthoryear{{Chevallier} \& {Polarski}}{{Chevallier} \&
  {Polarski}}{2001}]{2001IJMPD..10..213C}
{Chevallier} M.,  {Polarski} D.,  2001, \mn@doi [International Journal of
  Modern Physics D] {10.1142/S0218271801000822}, \href
  {https://ui.adsabs.harvard.edu/abs/2001IJMPD..10..213C} {10, 213}

\bibitem[\protect\citeauthoryear{{Child}, {Habib}, {Heitmann}, {Frontiere},
  {Finkel}, {Pope}  \& {Morozov}}{{Child} et~al.}{2018}]{2018ApJ...859...55C}
{Child} H.~L.,  {Habib} S.,  {Heitmann} K.,  {Frontiere} N.,  {Finkel} H.,
  {Pope} A.,   {Morozov} V.,  2018, \mn@doi [\apj] {10.3847/1538-4357/aabf95},
  \href {https://ui.adsabs.harvard.edu/abs/2018ApJ...859...55C} {859, 55}

\bibitem[\protect\citeauthoryear{{Chuang} et~al.,}{{Chuang}
  et~al.}{2019}]{2019MNRAS.487...48C}
{Chuang} C.-H.,  et~al., 2019, \mn@doi [\mnras] {10.1093/mnras/stz1233}, \href
  {https://ui.adsabs.harvard.edu/abs/2019MNRAS.487...48C} {487, 48}

\bibitem[\protect\citeauthoryear{{Contreras}, {Angulo}, {Zennaro}, {Aric{\`o}}
  \& {Pellejero-Iba{\~n}ez}}{{Contreras} et~al.}{2020}]{2020MNRAS.499.4905C}
{Contreras} S.,  {Angulo} R.~E.,  {Zennaro} M.,  {Aric{\`o}} G.,
  {Pellejero-Iba{\~n}ez} M.,  2020, \mn@doi [\mnras] {10.1093/mnras/staa3117},
  \href {https://ui.adsabs.harvard.edu/abs/2020MNRAS.499.4905C} {499, 4905}

\bibitem[\protect\citeauthoryear{{Correa}, {Wyithe}, {Schaye}  \&
  {Duffy}}{{Correa} et~al.}{2015}]{2015MNRAS.452.1217C}
{Correa} C.~A.,  {Wyithe} J. S.~B.,  {Schaye} J.,   {Duffy} A.~R.,  2015,
  \mn@doi [\mnras] {10.1093/mnras/stv1363}, \href
  {https://ui.adsabs.harvard.edu/abs/2015MNRAS.452.1217C} {452, 1217}

\bibitem[\protect\citeauthoryear{{Davis}, {Efstathiou}, {Frenk}  \&
  {White}}{{Davis} et~al.}{1985}]{1985ApJ...292..371D}
{Davis} M.,  {Efstathiou} G.,  {Frenk} C.~S.,   {White} S.~D.~M.,  1985,
  \mn@doi [\apj] {10.1086/163168}, \href
  {https://ui.adsabs.harvard.edu/abs/1985ApJ...292..371D} {292, 371}

\bibitem[\protect\citeauthoryear{Despali, Vegetti, White, Giocoli  \& van~den
  Bosch}{Despali et~al.}{2018}]{10.1093/mnras/sty159}
Despali G.,  Vegetti S.,  White S. D.~M.,  Giocoli C.,   van~den Bosch F.~C.,
  2018, \mn@doi [Monthly Notices of the Royal Astronomical Society]
  {10.1093/mnras/sty159}, 475, 5424

\bibitem[\protect\citeauthoryear{{Diemer} \& {Joyce}}{{Diemer} \&
  {Joyce}}{2019}]{2019ApJ...871..168D}
{Diemer} B.,  {Joyce} M.,  2019, \mn@doi [\apj] {10.3847/1538-4357/aafad6},
  \href {https://ui.adsabs.harvard.edu/abs/2019ApJ...871..168D} {871, 168}

\bibitem[\protect\citeauthoryear{{Dutton} \& {Macci{\`o}}}{{Dutton} \&
  {Macci{\`o}}}{2014}]{2014MNRAS.441.3359D}
{Dutton} A.~A.,  {Macci{\`o}} A.~V.,  2014, \mn@doi [\mnras]
  {10.1093/mnras/stu742}, \href
  {https://ui.adsabs.harvard.edu/abs/2014MNRAS.441.3359D} {441, 3359}

\bibitem[\protect\citeauthoryear{{Einasto}}{{Einasto}}{1965}]{1965TrAlm...5...87E}
{Einasto} J.,  1965, Trudy Astrofizicheskogo Instituta Alma-Ata, \href
  {https://ui.adsabs.harvard.edu/abs/1965TrAlm...5...87E} {5, 87}

\bibitem[\protect\citeauthoryear{{Fedeli}, {Bartelmann}, {Meneghetti}  \&
  {Moscardini}}{{Fedeli} et~al.}{2007}]{2007A&A...473..715F}
{Fedeli} C.,  {Bartelmann} M.,  {Meneghetti} M.,   {Moscardini} L.,  2007,
  \mn@doi [\aap] {10.1051/0004-6361:20077926}, \href
  {https://ui.adsabs.harvard.edu/abs/2007A&A...473..715F} {473, 715}

\bibitem[\protect\citeauthoryear{{Gao}, {Navarro}, {Cole}, {Frenk}, {White},
  {Springel}, {Jenkins}  \& {Neto}}{{Gao} et~al.}{2008}]{2008MNRAS.387..536G}
{Gao} L.,  {Navarro} J.~F.,  {Cole} S.,  {Frenk} C.~S.,  {White} S. D.~M.,
  {Springel} V.,  {Jenkins} A.,   {Neto} A.~F.,  2008, \mn@doi [\mnras]
  {10.1111/j.1365-2966.2008.13277.x}, \href
  {https://ui.adsabs.harvard.edu/abs/2008MNRAS.387..536G} {387, 536}

\bibitem[\protect\citeauthoryear{{Hellwing}, {Cautun}, {Knebe}, {Juszkiewicz}
  \& {Knollmann}}{{Hellwing} et~al.}{2013}]{2013JCAP...10..012H}
{Hellwing} W.~A.,  {Cautun} M.,  {Knebe} A.,  {Juszkiewicz} R.,   {Knollmann}
  S.,  2013, \mn@doi [\jcap] {10.1088/1475-7516/2013/10/012}, \href
  {https://ui.adsabs.harvard.edu/abs/2013JCAP...10..012H} {2013, 012}

\bibitem[\protect\citeauthoryear{Hunter}{Hunter}{2007}]{Matplotlib-Hunter07}
Hunter J.~D.,  2007, \mn@doi [Computing in Science \& Engineering]
  {10.1109/MCSE.2007.55}, 9, 90

\bibitem[\protect\citeauthoryear{{Huss}, {Jain}  \& {Steinmetz}}{{Huss}
  et~al.}{1999}]{1999ApJ...517...64H}
{Huss} A.,  {Jain} B.,   {Steinmetz} M.,  1999, \mn@doi [\apj]
  {10.1086/307161}, \href
  {https://ui.adsabs.harvard.edu/abs/1999ApJ...517...64H} {517, 64}

\bibitem[\protect\citeauthoryear{{Knebe} et~al.,}{{Knebe}
  et~al.}{2021}]{2021arXiv210313088K}
{Knebe} A.,  et~al., 2021, arXiv e-prints, \href
  {https://ui.adsabs.harvard.edu/abs/2021arXiv210313088K} {p. arXiv:2103.13088}

\bibitem[\protect\citeauthoryear{{Knollmann}, {Power}  \& {Knebe}}{{Knollmann}
  et~al.}{2008}]{2008MNRAS.385..545K}
{Knollmann} S.~R.,  {Power} C.,   {Knebe} A.,  2008, \mn@doi [\mnras]
  {10.1111/j.1365-2966.2008.12857.x}, \href
  {https://ui.adsabs.harvard.edu/abs/2008MNRAS.385..545K} {385, 545}

\bibitem[\protect\citeauthoryear{{Lacey} \& {Cole}}{{Lacey} \&
  {Cole}}{1993}]{1993MNRAS.262..627L}
{Lacey} C.,  {Cole} S.,  1993, \mn@doi [\mnras] {10.1093/mnras/262.3.627},
  \href {https://ui.adsabs.harvard.edu/abs/1993MNRAS.262..627L} {262, 627}

\bibitem[\protect\citeauthoryear{{Linden} \& {Virey}}{{Linden} \&
  {Virey}}{2008}]{2008PhRvD..78b3526L}
{Linden} S.,  {Virey} J.-M.,  2008, \mn@doi [\prd]
  {10.1103/PhysRevD.78.023526}, \href
  {https://ui.adsabs.harvard.edu/abs/2008PhRvD..78b3526L} {78, 023526}

\bibitem[\protect\citeauthoryear{{Linder}}{{Linder}}{2003}]{2003PhRvL..90i1301L}
{Linder} E.~V.,  2003, \mn@doi [\prl] {10.1103/PhysRevLett.90.091301}, \href
  {https://ui.adsabs.harvard.edu/abs/2003PhRvL..90i1301L} {90, 091301}

\bibitem[\protect\citeauthoryear{{Lucie-Smith}, {Adhikari}  \&
  {Wechsler}}{{Lucie-Smith} et~al.}{2022}]{2022arXiv220504474L}
{Lucie-Smith} L.,  {Adhikari} S.,   {Wechsler} R.~H.,  2022, arXiv e-prints,
  \href {https://ui.adsabs.harvard.edu/abs/2022arXiv220504474L} {p.
  arXiv:2205.04474}

\bibitem[\protect\citeauthoryear{{Ludlow} \& {Angulo}}{{Ludlow} \&
  {Angulo}}{2017}]{2017MNRAS.465L..84L}
{Ludlow} A.~D.,  {Angulo} R.~E.,  2017, \mn@doi [\mnras]
  {10.1093/mnrasl/slw216}, \href
  {https://ui.adsabs.harvard.edu/abs/2017MNRAS.465L..84L} {465, L84}

\bibitem[\protect\citeauthoryear{{Ludlow}, {Navarro}, {White},
  {Boylan-Kolchin}, {Springel}, {Jenkins}  \& {Frenk}}{{Ludlow}
  et~al.}{2011}]{2011MNRAS.415.3895L}
{Ludlow} A.~D.,  {Navarro} J.~F.,  {White} S. D.~M.,  {Boylan-Kolchin} M.,
  {Springel} V.,  {Jenkins} A.,   {Frenk} C.~S.,  2011, \mn@doi [\mnras]
  {10.1111/j.1365-2966.2011.19008.x}, \href
  {https://ui.adsabs.harvard.edu/abs/2011MNRAS.415.3895L} {415, 3895}

\bibitem[\protect\citeauthoryear{{Ludlow}, {Navarro}, {Li}, {Angulo},
  {Boylan-Kolchin}  \& {Bett}}{{Ludlow} et~al.}{2012}]{2012MNRAS.427.1322L}
{Ludlow} A.~D.,  {Navarro} J.~F.,  {Li} M.,  {Angulo} R.~E.,  {Boylan-Kolchin}
  M.,   {Bett} P.~E.,  2012, \mn@doi [\mnras]
  {10.1111/j.1365-2966.2012.21892.x}, \href
  {https://ui.adsabs.harvard.edu/abs/2012MNRAS.427.1322L} {427, 1322}

\bibitem[\protect\citeauthoryear{{Ludlow} et~al.,}{{Ludlow}
  et~al.}{2013}]{2013MNRAS.432.1103L}
{Ludlow} A.~D.,  et~al., 2013, \mn@doi [\mnras] {10.1093/mnras/stt526}, \href
  {https://ui.adsabs.harvard.edu/abs/2013MNRAS.432.1103L} {432, 1103}

\bibitem[\protect\citeauthoryear{{Ludlow}, {Navarro}, {Angulo},
  {Boylan-Kolchin}, {Springel}, {Frenk}  \& {White}}{{Ludlow}
  et~al.}{2014a}]{2014MNRAS.441..378L}
{Ludlow} A.~D.,  {Navarro} J.~F.,  {Angulo} R.~E.,  {Boylan-Kolchin} M.,
  {Springel} V.,  {Frenk} C.,   {White} S. D.~M.,  2014a, \mn@doi [\mnras]
  {10.1093/mnras/stu483}, \href
  {https://ui.adsabs.harvard.edu/abs/2014MNRAS.441..378L} {441, 378}

\bibitem[\protect\citeauthoryear{Ludlow, Navarro, Angulo, Boylan-Kolchin,
  Springel, Frenk  \& White}{Ludlow et~al.}{2014b}]{Ludlow_2014}
Ludlow A.~D.,  Navarro J.~F.,  Angulo R.~E.,  Boylan-Kolchin M.,  Springel V.,
  Frenk C.,   White S. D.~M.,  2014b, \mn@doi [Monthly Notices of the Royal
  Astronomical Society] {10.1093/mnras/stu483}, 441, 378–388

\bibitem[\protect\citeauthoryear{{Ludlow}, {Bose}, {Angulo}, {Wang},
  {Hellwing}, {Navarro}, {Cole}  \& {Frenk}}{{Ludlow}
  et~al.}{2016}]{2016MNRAS.460.1214L}
{Ludlow} A.~D.,  {Bose} S.,  {Angulo} R.~E.,  {Wang} L.,  {Hellwing} W.~A.,
  {Navarro} J.~F.,  {Cole} S.,   {Frenk} C.~S.,  2016, \mn@doi [\mnras]
  {10.1093/mnras/stw1046}, \href
  {https://ui.adsabs.harvard.edu/abs/2016MNRAS.460.1214L} {460, 1214}

\bibitem[\protect\citeauthoryear{{Ludlow}, {Schaye}  \& {Bower}}{{Ludlow}
  et~al.}{2019}]{2019MNRAS.488.3663L}
{Ludlow} A.~D.,  {Schaye} J.,   {Bower} R.,  2019, \mn@doi [\mnras]
  {10.1093/mnras/stz1821}, \href
  {https://ui.adsabs.harvard.edu/abs/2019MNRAS.488.3663L} {488, 3663}

\bibitem[\protect\citeauthoryear{{Ludlow}, {Schaye}, {Schaller}  \&
  {Bower}}{{Ludlow} et~al.}{2020}]{2020MNRAS.493.2926L}
{Ludlow} A.~D.,  {Schaye} J.,  {Schaller} M.,   {Bower} R.,  2020, \mn@doi
  [\mnras] {10.1093/mnras/staa316}, \href
  {https://ui.adsabs.harvard.edu/abs/2020MNRAS.493.2926L} {493, 2926}

\bibitem[\protect\citeauthoryear{{Macci{\`o}}, {Dutton}  \& {van den
  Bosch}}{{Macci{\`o}} et~al.}{2008}]{2008MNRAS.391.1940M}
{Macci{\`o}} A.~V.,  {Dutton} A.~A.,   {van den Bosch} F.~C.,  2008, \mn@doi
  [\mnras] {10.1111/j.1365-2966.2008.14029.x}, \href
  {https://ui.adsabs.harvard.edu/abs/2008MNRAS.391.1940M} {391, 1940}

\bibitem[\protect\citeauthoryear{{Maion}, {Angulo}  \& {Zennaro}}{{Maion}
  et~al.}{2022}]{2022arXiv220403868M}
{Maion} F.,  {Angulo} R.~E.,   {Zennaro} M.,  2022, arXiv e-prints, \href
  {https://ui.adsabs.harvard.edu/abs/2022arXiv220403868M} {p. arXiv:2204.03868}

\bibitem[\protect\citeauthoryear{{Mandelbaum}, {Seljak}  \&
  {Hirata}}{{Mandelbaum} et~al.}{2008}]{2008JCAP...08..006M}
{Mandelbaum} R.,  {Seljak} U.,   {Hirata} C.~M.,  2008, \mn@doi [\jcap]
  {10.1088/1475-7516/2008/08/006}, \href
  {https://ui.adsabs.harvard.edu/abs/2008JCAP...08..006M} {2008, 006}

\bibitem[\protect\citeauthoryear{{Mandic}, {Bird}  \& {Cholis}}{{Mandic}
  et~al.}{2016}]{2016PhRvL.117t1102M}
{Mandic} V.,  {Bird} S.,   {Cholis} I.,  2016, \mn@doi [\prl]
  {10.1103/PhysRevLett.117.201102}, \href
  {https://ui.adsabs.harvard.edu/abs/2016PhRvL.117t1102M} {117, 201102}

\bibitem[\protect\citeauthoryear{{Navarro}, {Frenk}  \& {White}}{{Navarro}
  et~al.}{1996}]{NFW}
{Navarro} J.~F.,  {Frenk} C.~S.,   {White} S. D.~M.,  1996, \mn@doi [\apj]
  {10.1086/177173}, \href
  {https://ui.adsabs.harvard.edu/abs/1996ApJ...462..563N} {462, 563}

\bibitem[\protect\citeauthoryear{{Navarro}, {Frenk}  \& {White}}{{Navarro}
  et~al.}{1997}]{1997ApJ...490..493N}
{Navarro} J.~F.,  {Frenk} C.~S.,   {White} S. D.~M.,  1997, \mn@doi [\apj]
  {10.1086/304888}, \href
  {https://ui.adsabs.harvard.edu/abs/1997ApJ...490..493N} {490, 493}

\bibitem[\protect\citeauthoryear{{Navarro} et~al.,}{{Navarro}
  et~al.}{2004}]{2004MNRAS.349.1039N}
{Navarro} J.~F.,  et~al., 2004, \mn@doi [\mnras]
  {10.1111/j.1365-2966.2004.07586.x}, \href
  {https://ui.adsabs.harvard.edu/abs/2004MNRAS.349.1039N} {349, 1039}

\bibitem[\protect\citeauthoryear{{Neto} et~al.,}{{Neto}
  et~al.}{2007}]{2007MNRAS.381.1450N}
{Neto} A.~F.,  et~al., 2007, \mn@doi [\mnras]
  {10.1111/j.1365-2966.2007.12381.x}, \href
  {https://ui.adsabs.harvard.edu/abs/2007MNRAS.381.1450N} {381, 1450}

\bibitem[\protect\citeauthoryear{{Okoli}, {Taylor}  \& {Afshordi}}{{Okoli}
  et~al.}{2018}]{2018JCAP...08..019O}
{Okoli} C.,  {Taylor} J.~E.,   {Afshordi} N.,  2018, \mn@doi [\jcap]
  {10.1088/1475-7516/2018/08/019}, \href
  {https://ui.adsabs.harvard.edu/abs/2018JCAP...08..019O} {2018, 019}

\bibitem[\protect\citeauthoryear{{Peebles}}{{Peebles}}{1980}]{1980lssu.book.....P}
{Peebles} P.~J.~E.,  1980, {The large-scale structure of the universe}

\bibitem[\protect\citeauthoryear{{Planck Collaboration} et~al.,}{{Planck
  Collaboration} et~al.}{2020}]{2020A&A...641A...6P}
{Planck Collaboration} et~al., 2020, \mn@doi [\aap]
  {10.1051/0004-6361/201833910}, \href
  {https://ui.adsabs.harvard.edu/abs/2020A&A...641A...6P} {641, A6}

\bibitem[\protect\citeauthoryear{{Power}, {Navarro}, {Jenkins}, {Frenk},
  {White}, {Springel}, {Stadel}  \& {Quinn}}{{Power}
  et~al.}{2003}]{2003MNRAS.338...14P}
{Power} C.,  {Navarro} J.~F.,  {Jenkins} A.,  {Frenk} C.~S.,  {White} S.~D.~M.,
   {Springel} V.,  {Stadel} J.,   {Quinn} T.,  2003, \mn@doi [\mnras]
  {10.1046/j.1365-8711.2003.05925.x}, \href
  {https://ui.adsabs.harvard.edu/abs/2003MNRAS.338...14P} {338, 14}

\bibitem[\protect\citeauthoryear{{Prada}, {Klypin}, {Cuesta}, {Betancort-Rijo}
  \& {Primack}}{{Prada} et~al.}{2012}]{2012MNRAS.423.3018P}
{Prada} F.,  {Klypin} A.~A.,  {Cuesta} A.~J.,  {Betancort-Rijo} J.~E.,
  {Primack} J.,  2012, \mn@doi [\mnras] {10.1111/j.1365-2966.2012.21007.x},
  \href {https://ui.adsabs.harvard.edu/abs/2012MNRAS.423.3018P} {423, 3018}

\bibitem[\protect\citeauthoryear{{Richardson}, {St{\"u}cker}, {Angulo}  \&
  {Hahn}}{{Richardson} et~al.}{2022}]{2022MNRAS.511.6019R}
{Richardson} T.~R.~G.,  {St{\"u}cker} J.,  {Angulo} R.~E.,   {Hahn} O.,  2022,
  \mn@doi [\mnras] {10.1093/mnras/stac493}, \href
  {https://ui.adsabs.harvard.edu/abs/2022MNRAS.511.6019R} {511, 6019}

\bibitem[\protect\citeauthoryear{{Riess} et~al.,}{{Riess}
  et~al.}{2016}]{2016ApJ...826...56R}
{Riess} A.~G.,  et~al., 2016, \mn@doi [\apj] {10.3847/0004-637X/826/1/56},
  \href {https://ui.adsabs.harvard.edu/abs/2016ApJ...826...56R} {826, 56}

\bibitem[\protect\citeauthoryear{{S{\'a}nchez-Conde} \&
  {Prada}}{{S{\'a}nchez-Conde} \& {Prada}}{2014}]{2014MNRAS.442.2271S}
{S{\'a}nchez-Conde} M.~A.,  {Prada} F.,  2014, \mn@doi [\mnras]
  {10.1093/mnras/stu1014}, \href
  {https://ui.adsabs.harvard.edu/abs/2014MNRAS.442.2271S} {442, 2271}

\bibitem[\protect\citeauthoryear{{Springel}}{{Springel}}{2015}]{2015ascl.soft02003S}
{Springel} V.,  2015, {N-GenIC: Cosmological structure initial conditions}
  (\mn@eprint {ascl} {1502.003})

\bibitem[\protect\citeauthoryear{Springel, White, Tormen  \&
  Kauffmann}{Springel et~al.}{2001}]{10.1046/j.1365-8711.2001.04912.x}
Springel V.,  White S. D.~M.,  Tormen G.,   Kauffmann G.,  2001, \mn@doi
  [Monthly Notices of the Royal Astronomical Society]
  {10.1046/j.1365-8711.2001.04912.x}, 328, 726

\bibitem[\protect\citeauthoryear{{Springel} et~al.,}{{Springel}
  et~al.}{2008}]{2008MNRAS.391.1685S}
{Springel} V.,  et~al., 2008, \mn@doi [\mnras]
  {10.1111/j.1365-2966.2008.14066.x}, \href
  {https://ui.adsabs.harvard.edu/abs/2008MNRAS.391.1685S} {391, 1685}

\bibitem[\protect\citeauthoryear{{Wang} \& {White}}{{Wang} \&
  {White}}{2009}]{2009MNRAS.396..709W}
{Wang} J.,  {White} S. D.~M.,  2009, \mn@doi [\mnras]
  {10.1111/j.1365-2966.2009.14755.x}, \href
  {https://ui.adsabs.harvard.edu/abs/2009MNRAS.396..709W} {396, 709}

\bibitem[\protect\citeauthoryear{{Wang}, {Bose}, {Frenk}, {Gao}, {Jenkins},
  {Springel}  \& {White}}{{Wang} et~al.}{2020}]{2020Natur.585...39W}
{Wang} J.,  {Bose} S.,  {Frenk} C.~S.,  {Gao} L.,  {Jenkins} A.,  {Springel}
  V.,   {White} S.~D.~M.,  2020, \mn@doi [\nat] {10.1038/s41586-020-2642-9},
  \href {https://ui.adsabs.harvard.edu/abs/2020Natur.585...39W} {585, 39}

\bibitem[\protect\citeauthoryear{{Wechsler}, {Bullock}, {Primack}, {Kravtsov}
  \& {Dekel}}{{Wechsler} et~al.}{2002}]{2002ApJ...568...52W}
{Wechsler} R.~H.,  {Bullock} J.~S.,  {Primack} J.~R.,  {Kravtsov} A.~V.,
  {Dekel} A.,  2002, \mn@doi [\apj] {10.1086/338765}, \href
  {https://ui.adsabs.harvard.edu/abs/2002ApJ...568...52W} {568, 52}

\bibitem[\protect\citeauthoryear{{Zennaro}, {Bel}, {Villaescusa-Navarro},
  {Carbone}, {Sefusatti}  \& {Guzzo}}{{Zennaro}
  et~al.}{2017}]{2017MNRAS.466.3244Z}
{Zennaro} M.,  {Bel} J.,  {Villaescusa-Navarro} F.,  {Carbone} C.,  {Sefusatti}
  E.,   {Guzzo} L.,  2017, \mn@doi [\mnras] {10.1093/mnras/stw3340}, \href
  {https://ui.adsabs.harvard.edu/abs/2017MNRAS.466.3244Z} {466, 3244}

\bibitem[\protect\citeauthoryear{{Zennaro}, {Angulo}, {Aric{\`o}}, {Contreras}
  \& {Pellejero-Ib{\'a}{\~n}ez}}{{Zennaro} et~al.}{2019}]{2019MNRAS.489.5938Z}
{Zennaro} M.,  {Angulo} R.~E.,  {Aric{\`o}} G.,  {Contreras} S.,
  {Pellejero-Ib{\'a}{\~n}ez} M.,  2019, \mn@doi [\mnras]
  {10.1093/mnras/stz2612}, \href
  {https://ui.adsabs.harvard.edu/abs/2019MNRAS.489.5938Z} {489, 5938}

\bibitem[\protect\citeauthoryear{{Zhang}, {Liao}, {Li}  \& {Gao}}{{Zhang}
  et~al.}{2019}]{2019MNRAS.487.1227Z}
{Zhang} T.,  {Liao} S.,  {Li} M.,   {Gao} L.,  2019, \mn@doi [\mnras]
  {10.1093/mnras/stz1370}, \href
  {https://ui.adsabs.harvard.edu/abs/2019MNRAS.487.1227Z} {487, 1227}

\bibitem[\protect\citeauthoryear{{Zhao}, {Jing}, {Mo}  \& {B{\"o}rner}}{{Zhao}
  et~al.}{2003}]{2003ApJ...597L...9Z}
{Zhao} D.~H.,  {Jing} Y.~P.,  {Mo} H.~J.,   {B{\"o}rner} G.,  2003, \mn@doi
  [\apjl] {10.1086/379734}, \href
  {https://ui.adsabs.harvard.edu/abs/2003ApJ...597L...9Z} {597, L9}

\bibitem[\protect\citeauthoryear{{van der Walt}, {Colbert}  \&
  {Varoquaux}}{{van der Walt} et~al.}{2011}]{Numpy-vanDerWalt11}
{van der Walt} S.,  {Colbert} S.~C.,   {Varoquaux} G.,  2011, \mn@doi
  [Computing in Science Engineering] {10.1109/MCSE.2011.37}, 13, 22

\makeatother
\end{thebibliography}
